 \documentclass[final,authoryear,1p,11pt]{elsarticle}

\usepackage{graphicx}
\usepackage{epstopdf}
\usepackage{booktabs}
\usepackage{natbib}
\usepackage{fancyhdr}
\usepackage{amsmath}
\usepackage{amsfonts}
\usepackage{multirow}
\usepackage{lscape}
\usepackage{color}
\usepackage{hyperref}
\usepackage{color,soul}
\sethlcolor{green}
\usepackage{ifthen}
\usepackage[latin1]{inputenc}
\usepackage{titletoc}
\usepackage{eso-pic}
\definecolor{123}{rgb}{.9,.9,.9}
\hypersetup{
colorlinks=false,
citecolor=blue,
linkbordercolor={1 1 1}, 
citebordercolor={1 1 1},
urlbordercolor={1 1 1}
}

\usepackage{amssymb}
 \usepackage{amsthm}
 \usepackage{xcolor}

\newcommand{\ra}[1]{\renewcommand{\arraystretch}{#1}}

\begin{document}

\begin{frontmatter}

%
%
\title{Understanding the source of multifractality in financial markets \tnoteref{label1} }
%
%
%
\author[ies,utia]{Jozef Barunik\corref{cor2} } \ead{barunik@utia.cas.cz}
\author[kent]{Tomaso Aste}
\author[kings]{Tiziana Di Matteo} 
\author[austr]{Ruipeng Liu}
\cortext[cor2]{Corresponding author}
\address[ies]{Institute of Economic Studies, Charles University, Opletalova 21, 110 00, Prague,  CR}
\address[utia]{Institute of Information Theory and Automation, Academy of Sciences of the Czech Republic, Pod Vodarenskou Vezi 4, 182 00, Prague, Czech Republic}
\address[kent]{School of Physical Sciences, University of Kent, United Kingdom and The Australian National University, Australia}
\address[kings]{Department of Mathematics, King's College London, Strand, London WC2R 2LS, United Kingdom}
\address[austr]{School of Accounting, Economics \& Finance, Deakin University, 221 Burwood Highway, Melbourne, VIC 3125, Australia}
%
\begin{abstract}
In this paper, we use the generalized Hurst exponent approach to study the multi-scaling behavior of different financial time series.
We  show that this approach is robust and powerful in detecting different types of multiscaling.
We observe a puzzling phenomenon where an apparent increase in multifractality is measured in time series generated from shuffled returns, where all time-correlations are destroyed, while the return distributions are conserved.
This effect is robust and it is reproduced in several real financial data including stock market indices, exchange rates and interest rates.
In order to understand the origin of this effect we investigate different simulated time series by means of the Markov switching multifractal (MSM) model, autoregressive fractionally integrated moving average (ARFIMA) processes with stable innovations, fractional Brownian motion and Levy flights. 
Overall we conclude that the multifractality observed in financial time series is mainly a consequence of the characteristic fat-tailed distribution of the returns and  time-correlations have the effect to decrease the measured multifractality.
\end{abstract}
\end{frontmatter}


\section{Introduction}
A traditional assumption, used in the early studies of financial time series, considered that returns are independent, Gaussian random variables.
However, uncountable number of empirical studies, initiated by \cite{mandelbrot}, have shown that empirical returns reveal instead very rich and non trivial statistical features, such as fat tails, volatility clustering and multiscaling. 
From that times, several models have been proposed to mimic the multiscaling behavior of stock market returns.
For instance, Benoit Mandelbrot, together with his students Luarent Calvet and Adlai Fisher, introduced a stochastic process as generating mechanism of stock market returns with a multifractal cascade \citep{mandelbrotcalvetfisher,mandelbrot99,calvetfisher}.
Such multifractal processes provide us with a new model with attractive stochastic properties, which can reproduce some stylized facts of financial markets: fat tails, volatility clustering, long-term dependence and multi-scaling.
However, the practical applicability of earlier versions of multifractal models suffers from its combinatorial nature and from its non-stationarity due to the restriction to a bounded interval.
The most attractive feature of these processes is their ability to generate several degrees of long memory in different powers of returns.
More recently  \cite{calvetfisher} and \cite{calvetfisher2004} introduced a new family of iterative multifractal models: the Markov-switching multifractal (MSM) model which preserves the hierarchical, multiplicative structure of the earlier models, but possesses appealing asymptotic properties.

In the recent years, there has been an increasing interest in the application of the scaling concept to financial markets \cite{muller90,Lux99,Dacorogna01a,Lux04a,carbone2004time}.
Scaling properties in time series have been studied in the literature by using several techniques.
For the interested reader let us briefly mention here some of them such as the seminal work of \cite{Hurst51} on rescaled range statistical analysis and the modified rescaled range analysis of \cite{Lo91}, the multi-affine analysis \citep{Peng94}, the detrended fluctuation analysis \citep{Ausloos,Bartolozzi07,DiMatteo07}, or its generalization, multifractal detrended fluctuation analysis \citep{Kantelhardt2002}.
The challenge for empirical and theoretical researches lies in uncovering what  scaling laws tell us about the underling mechanisms that generate the data.
Furthermore,  the empirical scaling evidences should be used as stylized facts that any theoretical model should also reproduce.

 In addition to this findings, \cite{schmittschertzer} show that the additive models like Brownian, fractional Brownian, L\'evy, Truncated L\'evy and fractional L\'evy models are not compatible with the properties of financial data and they propose the multifractal framework as an alternative. \cite{bianchi} argue that partition function of generally non-multifractal processes fitted to the financial time series behaves as those of a genuine multifractal process.
\cite{Jiang} find that scaling behavior of the original financial datasets can not be distinguished from those of shuffled time series.
\cite{Zhou} investigates the components of the empirical multifractality of financial returns and finds that temporal structure has minor impact on the multifractal spectrum.
More recently, \cite{schmitt} find that introduction of Euro had no influence on the statistical properties of the fluctuations of the Euro-Yuan exchange rate.


In this paper, we analyze the multi-scaling properties of different time series by means of the generalized Hurst exponent (GHE) which provides a robust estimator to compute these scaling properties \citep{DiMatteo03,DiMatteo05}.
There are two types of scaling behaviors studied in the finance literature: the behavior of the returns distribution tails as a function of the movement size, but keeping the time interval of the returns constant; the behavior of some forms of volatility measure as a function of the time interval on which the returns are measured.
In this study we investigate the link between the two in real and simulated data series.
Furthermore, to distinguish between the effects on multifractality from time-correlations and from fat-tailed return distributions we apply the GHE on shuffled  data series where the time history is destroyed but the return distribution is maintained.

The main part of this paper concerns the study of the source of the multifractality in financial datasets.
\cite{Kantelhardt2002} point out that in general, we can find two types of multifractality in the time series:
(i) Multifractality due to a broad probability density function;
(ii) Multifractality due to different long-range correlations of the small and large fluctuations in time.
In the first case, multifractality can not be removed by shuffling the series.
In the second case, the corresponding shuffled series should exhibit uni-fractality, since all long-range correlations are destroyed by shuffling.
In case that both types are present in the data, shuffled data should show different multifractality than the original series.

We contribute to the debate about scaling properties of the financial returns with a rigorous statistical analysis of the problem.
In particular, we investigate the two types of multifractality both on real financial data and MSM simulated time series.
To test the robustness of our findings, we also compare the results to the simulations from $\alpha$-stable distribution, fractional Brownian motion and  autoregressive fractionally integrated moving average model with stable innovations which allows us to study the impact of short memory in the heavy tailed process with long range dependence.
Our study is structured as follows.
Sections \ref{MSM} and \ref{sectghe} review the Markov-switching multifractal (MSM) and the generalized Hurts exponent (GHE) methods.
Section 4  reports the empirical and simulation-based results describing the two types of multifractality in the data.
In Section 5 we check the results for robustness by comparing the simulations from $\alpha$-stable distribution, fractional Brownian motion and fractional autoregressive moving average model with stable innovations.
Finally, results are followed by the conclusions given in Section 6.

\section{Markov-switching multifractal model}
\label{MSM}

In the Markov-switching multifractal model \citep{calvetfisher2004,LiuAste08,Lux2007,Liu07,Liu08} asset returns are modeled as:
\begin{equation}
\label{returns}
r_t=\sigma_t  u_t
\end{equation}
with innovations drown from a normal distribution with average zero and unitary standard deviation ($u_t \sim N(0,1)$) and instantaneous volatility determined by the product of $k$ volatility components, or multipliers, $M_t^{(1)},M_t^{(2)},\dots,M_t^{(k)}$ and a constant scale factor $\sigma$:
\begin{equation}
\sigma_t^2=\sigma^2 \prod_{i=1}^k M_t^{(i)},
\end{equation}
where $M_t^{(i)}$ is a random variable drawn from a binomial
distribution, which is characterized by random draws taking two
discrete values with equal probability, i.e.,  $m_0$ and $m_1$,
(with $m_1 = 2- m_0$, and  $1 \leq m_0 \leq 2$). Each volatility
component $M_t^{(i)}$  is renewed at time $t$ with probability
$\gamma_i$ depending on its rank $i$ within the hierarchy of
multipliers and it remains unchanged with probability $1-\gamma_i$.
The transition probabilities are:
\begin{equation}
\label{gamma}
\gamma_i=1-\left(1-\gamma_k\right)^{b^{i-k}}, \hspace{1cm} i=1,\dots,k,
\end{equation}
with  $\gamma_k \in\left[0,1\right]$ and $b\in\left(1,\infty\right)$. Different specifications of Eq. \ref{gamma} can be arbitrarily imposed, i.e. as in \cite{Lux2007}. By fixing $b=2$ and $\gamma_k=0.5$, we arrive to a parsimonious specification:
\begin{equation}
\label{gamma1}
\gamma_i=1-0.5^{2^{i-k}}, \hspace{1cm} i=1,\dots,k.
\end{equation}

The dynamics resulting from Eq. \ref{returns} is a particular version of a stochastic volatility model. 
With its rather parsimonious approach, this multi-fractal process preserves the hierarchical structure of  the original multi-fractal model of asset returns while dispensing with its restriction to a bounded interval.

\section{Generalized Hurst exponent}
\label{sectghe}
The generalized Hurst exponent (GHE) method is used to estimate the scaling exponents and study the multifractality of the empirical data.
The biggest advantage of GHE is that it combines the sensitivity to any type of dependence in the data to a computationally straightforward and simple algorithm\footnote{The algorithm provided by T. Aste can be freely downloaded at \url{http://www.mathworks.com/matlabcentral/fileexchange/30076}.}.

The generalized Hurst exponent method aims to extend the traditional scaling exponent methodology, providing a natural, unbiased, statistically and computationally efficient estimator able to capture the scaling features of financial fluctuations \citep{DiMatteo03,DiMatteo05,DiMatteo07,morales2011dynamical}. It is essentially a tool to study directly the scaling properties of the data via the $q-$th order moments of the distribution of the increments. Different exponents $q$ are associated with different characterizations of the multi-scaling behavior of the signal $X(t)$.

We consider the $q$-order moment of the distribution of the increments (\cite{barabasi}) of a time series $X(t)$:
\begin{equation}
\label{K_q}
K_q(\tau) = \frac{\langle \mid X(t+\tau) - X(t) \mid^q\rangle}{\langle \mid
X(t)\mid^q\rangle},
\end{equation}

where  $\tau$ varies in the interval between day and $\tau_{max}$ days. The generalized Hurst exponent $H(q)$ is then defined from the scaling behavior of $K_q(\tau)$, which can be assumed ((\cite{barabasi})) to follow the relation:
\begin{equation}
\label{K_qtau}
K_q(\tau) \sim {\left(\tau \right )}^{qH(q)}.
\end{equation}

We would like to remind the reader that this framework is based on the assumption that the process under study has such a scaling property and  the scaling property  stays unchanged across the observation time window.
In practice, financial time series show evidence of variation of their statistical properties with time, and show dependencies on the observation window.
The simplest case might be a linear drift $\eta t$ added to a stochastic variable $X(t)=\widetilde{X}(t)+\eta t$ with $\widetilde{X}(t)$ satisfying Eq. \ref{K_qtau}.
To apply scaling analysis to $\widetilde{X}(t)$ process, we need to subtract the drift $\eta t$ from the variable $X(t)$ where $\eta$ can be evaluated from the $\langle X(t+\tau)-X(t)\rangle=\eta \tau$.
This may be viewed as a first-order approximation, crucial for the correctness of the results, and such a linear detrending is performed in our computations.

For some values of $q$, the exponents are associated with special features. For instance $H(1)$ describes the scaling behavior of the absolute values of the increments. The value of this exponent is expected to be closely related to the original Hurst exponent, H, that is indeed associated with the scaling of the absolute spread in the increments. $H(2)$ is instead associated with the scaling of the autocorrelation function.


\section{Distinguishing between two types of multifractality}
\label{secdata}

\subsection{Data description}
We consider daily data for a collection of stock exchange indices: the Dow Jones Composite 65 Average Index ($Dow$) and $NIKKEI$ 225 Average Index ($Nik$).
Foreign exchange rates: British Pound to U.S. Dollar ($US/UK$), German Mark to U.S. Dollar ($DM/US$) and U.S. treasury bond rates with maturity 1 year and 2 years, 3 years, 5 years and 10 years  ($TB1$, $TB2$, $TB3$, $TB5$ and $TB10$ respectively) all over the period from June 1976 to November 2010. 
The daily prices are denoted as $p_t$, and returns are calculated as $r_t = \ln(p_t)-\ln(p_{t-1})$ for stock indices and foreign exchange rates and as $r_t = p_t - p_{t-1}$ for $TB1$, $TB2$, $TB3$, $TB5$ and $TB10$.

\subsection{MSM estimation of the parameters}

As  first step, the parameters $m_0$ and $\sigma$ in the MSM model are estimated, for the following hypothetical number of cascades in volatility $k=\left\{5,10,15,20\right\}$, by following the GMM (Generalized Method of Moments) approach proposed by \cite{Lux2007} using the same analytical moments as in his paper. 
Table \ref{gmm} reports the results, where the numbers within the parentheses are the standard errors.
We observe that the results for $k>10$ are almost identical. 
In fact, analytical moment conditions in \cite{Lux2007} show that higher cascade levels make smaller and smaller contributions to the moments so that their numerical values would stay almost constant.
If one monitors the development of estimated parameters with increasing $k$, one finds strong variations initially with a pronounced decrease of the estimates which become slower and slower until, eventually a constant value is reached somewhere around $k=10$ depending on individual time series.
Based on these estimated parameters, we have proceeded with an analysis of simulated data. 

\begin{table}
\scriptsize \caption{GMM estimations of MSM model for different $k$.
This table shows empirical GMM estimates of MSM model with
different $k$ and numbers in parenthesis are standard errors; All
empirical returns are demeaned prior to estimation.} \label{gmm}
\centering
\begin{tabular}{lrrrrrrrrrrr}
\toprule
&\multicolumn{2}{c}{$k$ = 5}& &\multicolumn{2}{c}{$k$ =10}& & \multicolumn{2}{c}{$k$ =15}& &\multicolumn{2}{c}{$k$ = 20}\\
\cmidrule{2-3} \cmidrule{5-6} \cmidrule{8-9} \cmidrule{11-12}
&  $\hat{m}_0$ & $\hat{\sigma}$& & $\hat{m}_0$ & $\hat{\sigma}$ & & $\hat{m}_0$ & $\hat{\sigma}$ & &$\hat{m}_0$ & $\hat{\sigma}$ \\
\cmidrule{2-3} \cmidrule{5-6} \cmidrule{8-9} \cmidrule{11-12}
$Dow$ &1.362 &0.016 & & 1.317 &0.013  & & 1.318 &0.012  & & 1.318 &0.011    \\
       &(0.073)& (0.003)  & & (0.091)  &(0.004)  & & (0.053)&(0.004)& & (0.055)  &(0.003)\\
$Nik$   & 1.449 &0.014 & & 1.437 &0.012 & & 1.434 &0.010 & & 1.434 &0.010  \\
         &(0.047)& (0.003) & & (0.050)&(0.003)& & (0.051) &(0.002) & &(0.051)&(0.002)\\
$DM/US$   & 1.257 &0.017 & & 1.205 &0.015 & & 1.203 &0.014 & & 1.203 &0.014 \\
         &(0.029)& (0.003) & & (0.031)&(0.003)& & (0.032) &(0.004)& & (0.032)&(0.003)\\
$US/UK$  &1.417 &0.017 & & 1.358 &0.011 & & 1.357 &0.011 & & 1.357 &0.010 \\
         &(0.033)& (0.004) & & (0.031)&(0.003)& & (0.032) &(0.004)& & (0.032)&(0.002)\\
$TB1$&1.630 &0.116 & & 1.596 &0.118 & & 1.595 &0.119 & & 1.595 &0.119\\
        &(0.042)&(0.010)  & & (0.041) &(0.013)& & (0.042)&(0.012)& & (0.042)&(0.013)\\
$TB2$ &1.707 &0.098 & & 1.671& 0.101  & &1.669 &0.101 & & 1.669 &0.102\\
      &(0.023)&(0.017) & & (0.023) &(0.016)& & (0.024)&(0.017)& & ( 0.023)&(0.015)\\
$TB3$ &1.616 &0.097 & & 1.588 &0.098 & & 1.586 &0.098 & & 1.586 &0.099\\
       &(0.020)&(0.012) & & (0.023) &(0.013)& & (0.024)&(0.013)& & (0.021)&(0.013)\\
$TB5$ &1.655 &0.088 & & 1.620 &0.090 & & 1.619 &0.091 & & 1.618 &0.091\\
       &(0.030)& (0.015)  & & (0.028)  &(0.014)  & & (0.027)&(0.014)& & (0.028)  &(0.013)\\
$TB10$ &1.690 &0.076 & & 1.664 &0.078 & & 1.662 &0.079& &  1.662& 0.078\\
       &(0.022)&(0.013) & & (0.023) &(0.012)& & (0.023)&(0.012)& & (0.023)&(0.010)\\
\bottomrule
\end{tabular}
\end{table}

\subsection{MSM simulation}

For each stock market index, for each set of estimated parameters ($\hat m_0$, $\hat \sigma$) and for each $k$, we have simulated 1000 MSM paths and computed the $H(q)$ with $q=\{1,2,3\}$ using GHE for empirical series as well as the simulated ones. 
First we verify that Eq. \ref{K_qtau} holds for $\tau \in[1,\tau_{max}]$.
Then we compute the values of $H(q)$ as averages of the best fit exponents corresponding to different $\tau_{max} \in \left[5,19\right]$. 
The whole procedure is repeated for different stochastic variables $X(t)$ from Eq. \ref{K_q}, namely $X(t) = p_{t}$, $X(t) = \sum_{t' = 1}^t |r_{t'}|$  and $X(t) = \sum_{t' = 1}^t r_{t'}^2$ representing prices and volatility estimates.

\begin{figure}
   \centering
   \includegraphics[width=\textwidth]{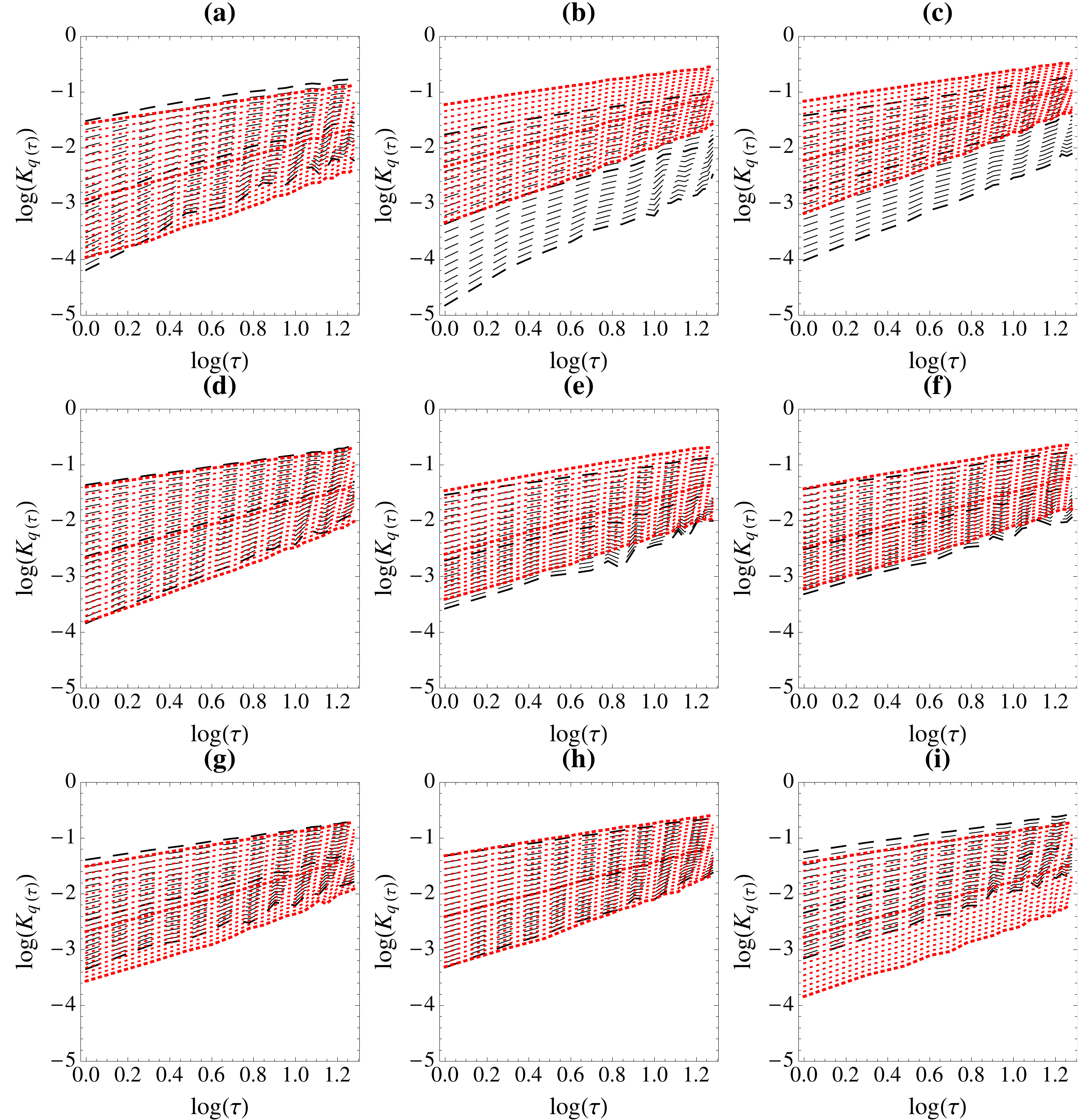}
   \caption{(Color online) Empirical $K_q(\tau)$ (and shuffled data in red) as a function of $\tau=1,\dots,19$ days on a log-log scale for $X(t)=p_t$. Each curve corresponds to different fixed values of $q=1,\dots,3$ from upper to bottom line, while upper bold line, middle bold line and bottom bold line correspond $q=1$, $q=2$ and $q=3$ respectively. (a) Dow (b) Nik (c) DM/US (d) US/UK (e) TB1 (f) TB2 (g) TB3 (h) TB5 (i) TB10. Note that axes are fixed for all plots which make them comparable.}
   \label{plotK}
\end{figure}

\begin{figure}
   \centering
   \includegraphics[width=\textwidth]{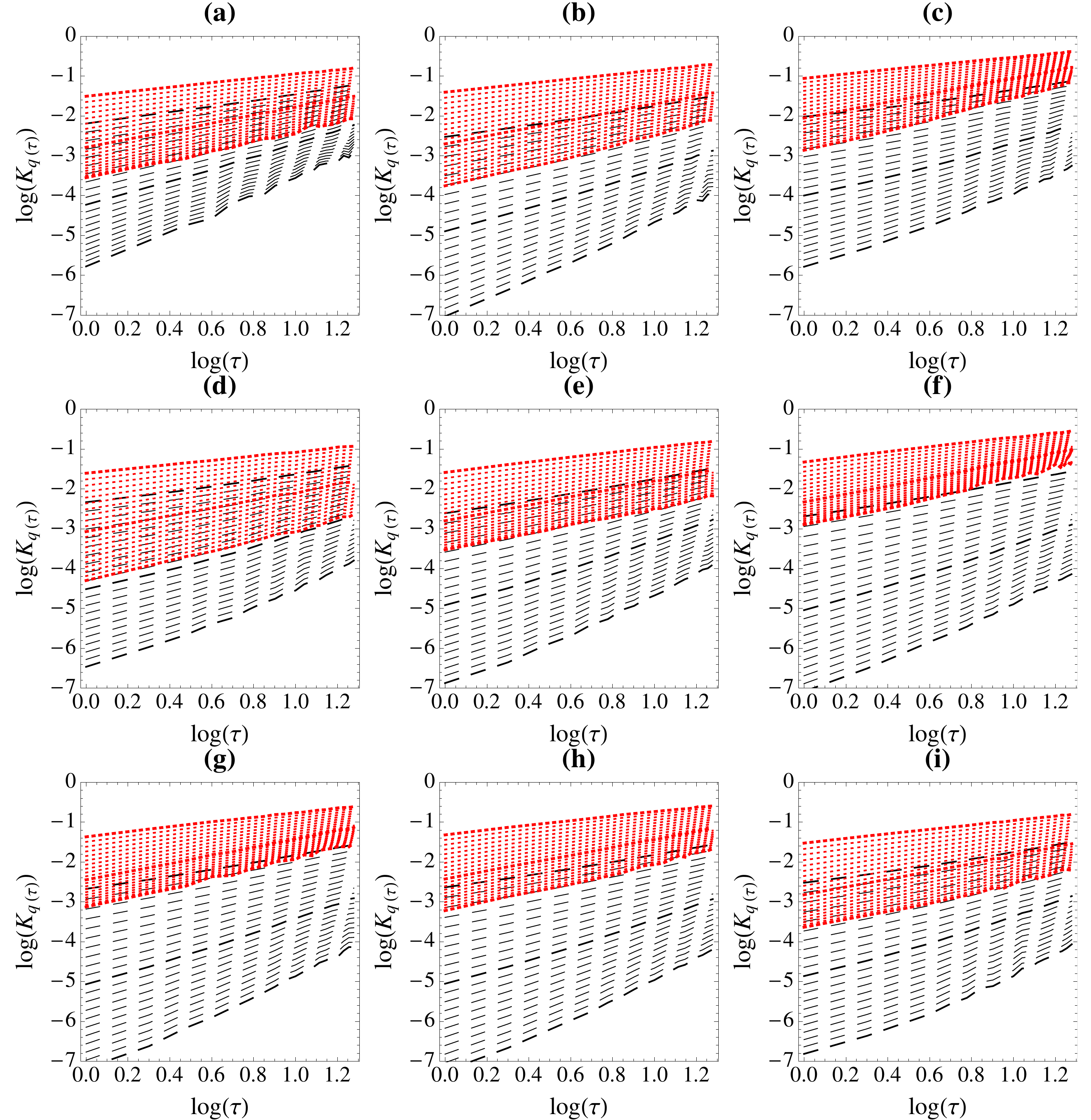}
   \caption{(Color online) Empirical $K_q(\tau)$ (and shuffled data in red) as a function of $\tau=1,\dots,19$ days on a log-log scale for $X(t)=\sum |r_{t'}|$. Each curve corresponds to different fixed values of $q=1,\dots,3$ from upper to bottom line, while upper bold line, middle bold line and bottom bold line correspond $q=1$, $q=2$ and $q=3$ respectively. (a) Dow (b) Nik (c) DM/US (d) US/UK (e) TB1 (f) TB2 (g) TB3 (h) TB5 (i) TB10. Note that axes are fixed for all plots which make them comparable.}
   \label{plotKabs}
\end{figure}

\begin{figure}
   \centering
   \includegraphics[width=\textwidth]{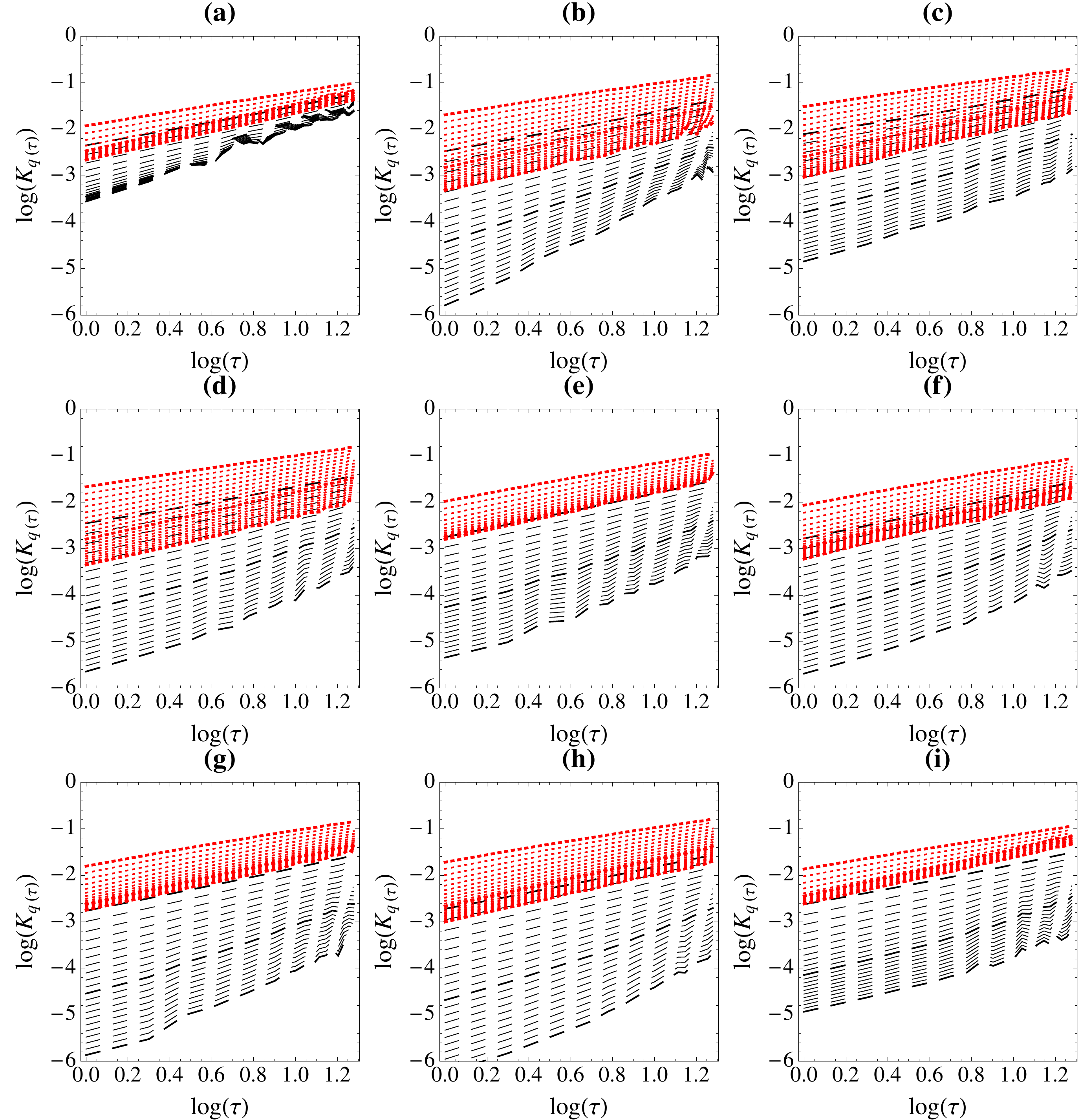}
   \caption{(Color online) Empirical $K_q(\tau)$ (and shuffled data in red) as a function of $\tau=1,\dots,19$ days on a log-log scale for $X(t)=\sum r_{t'}^2$. Each curve corresponds to different fixed values of $q=1,\dots,3$ from upper to bottom line, while upper bold line, middle bold line and bottom bold line correspond $q=1$, $q=2$ and $q=3$ respectively.  (a) Dow (b) Nik (c) DM/US (d) US/UK (e) TB1 (f) TB2 (g) TB3 (h) TB5 (i) TB10. Note that axes are fixed for all plots which make them comparable.}
   \label{plotKsqr}
\end{figure}

\begin{figure}
   \centering
   \includegraphics[width=\textwidth]{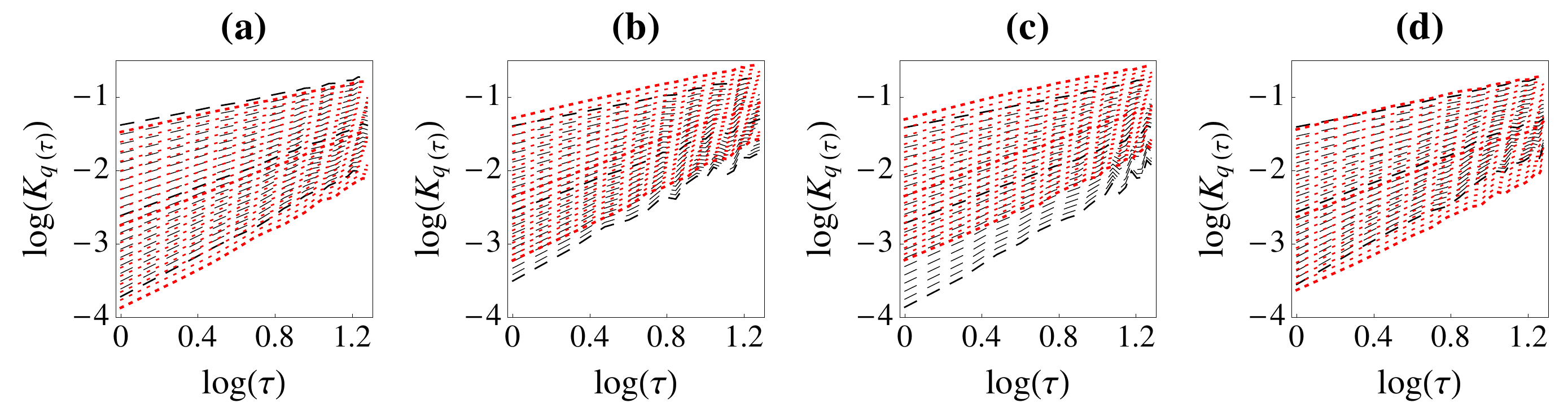}
   \caption{(Color online) Example of $K_q(\tau)$ (and shuffled data in red) on MSM simulated data (Nikkei estimates) as a function of $\tau=1,\dots,19$ days on a log-log scale for $X(t)=p_t$. Each curve corresponds to different fixed values of $q=1,\dots,3$ from upper to bottom line, while upper red line, middle red line and bottom line correspond $q=1$, $q=2$ and $q=3$ respectively. Simulated data for (a) $k=5$ (b) $k=10$ (c) $k=15$ (d) $k=20$.}
   \label{plotMSM1}
\end{figure}

\begin{figure}
   \centering
   \includegraphics[width=\textwidth]{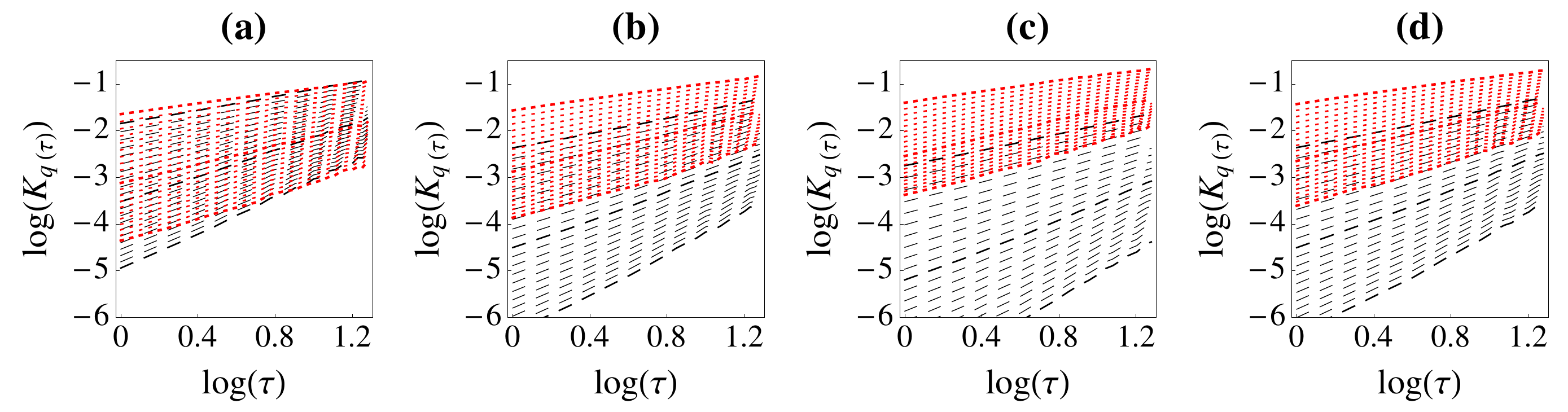}
   \caption{(Color online) Example of $K_q(\tau)$ (and shuffled data in red) on MSM simulated data (Nikkei estimates) as a function of $\tau=1,\dots,19$ days on a log-log scale for $X(t)=\sum |r_{t'}|$. Each curve corresponds to different fixed values of $q=1,\dots,3$ from upper to bottom line, while upper bold line, middle bold line and bottom bold line correspond $q=1$, $q=2$ and $q=3$ respectively. Simulated data for (a) $k=5$(b) $k=10$ (c) $k=15$ (d) $k=20$.}
   \label{plotMSM2}
\end{figure}

\begin{figure}
   \centering
   \includegraphics[width=\textwidth]{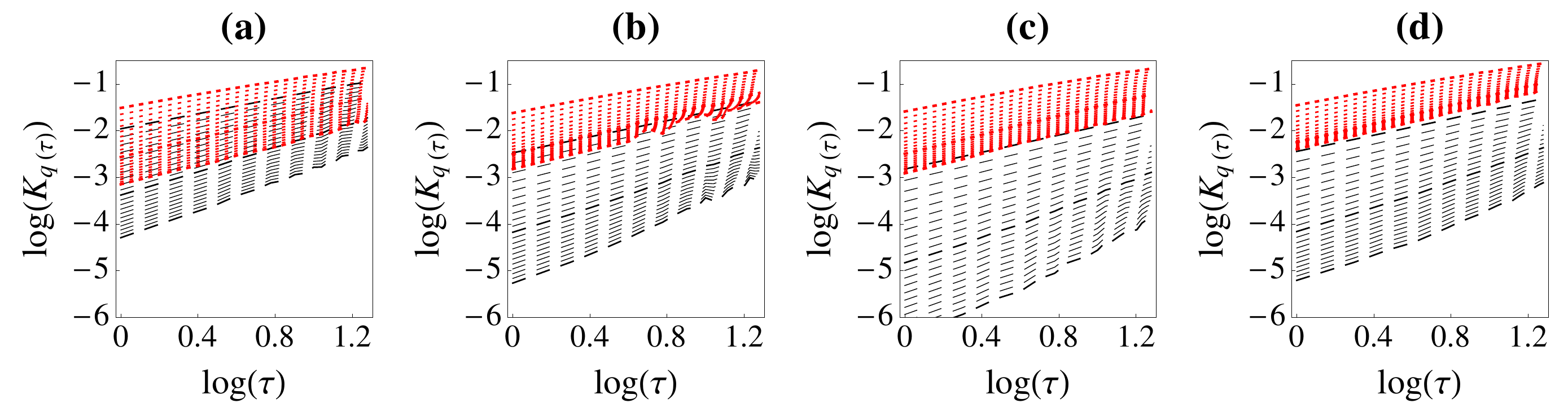}
   \caption{(Color online) Example of $K_q(\tau)$ (and shuffled data in red) on MSM simulated data (Nikkei estimates) as a function of $\tau=1,\dots,19$ days on a log-log scale $X(t)=\sum r_{t'}^2$. Each curve corresponds to different fixed values of $q=1,\dots,3$ from upper to bottom line, while upper bold line, middle bold line and bottom bold line correspond $q=1$, $q=2$ and $q=3$ respectively. Simulated data for (a) $k=5$(b) $k=10$ (c) $k=15$ (d) $k=20$.}
   \label{plotMSM3}
\end{figure}

We compute the $99\%$ confidence intervals of all exponents using different $\tau_{max}$ values and find that they are stable in their values within a range of $10\%$. Figures \ref{plotK}, \ref{plotKabs} and \ref{plotKsqr} show the $K_q(\tau)$ curves for $q=[1,\dots,3]$ on empirical data, while Figures \ref{plotMSM1}, \ref{plotMSM2} and \ref{plotMSM3} show examples of scaling of MSM simulated data on the Nikkei estimates\footnote{Similar scaling behaviors have been also found for the other simulated time series based on estimated coefficients from Table \ref{gmm} and for different considered stochastic variables.}. We report all results for different stochastic variables $p_t$, $\sum |r_{t'}|$ and $\sum r_{t'}^2$ and we can see that scaling holds for all tested series and thus it is appropriate to compute the Hurst exponent from the relation in Eq. \ref{K_qtau}.

\subsection{Multi-scaling of the data}

When studying the scaling exponents for different $q$, one can distinguish between two kind of processes: (i) a process where $H(q)=H$, constant, independent of $q$; (ii) a process with $H(q)$ not constant.
The first case characterizes uni-scaling or uni-fractality and its scaling behavior is determined from a unique constant value of the Hurst exponent: $H$. 
For instance, this might be the case of the self-affine process where $qH(q)$ is linear and fully determined by its $H$, for example Brownian motion. The second case, when $H(q)$ depends on $q$, the process is commonly called multi-scaling, or multi-fractal and different exponents characterize the scaling of different $q$ moments of the distribution. \cite{DiMatteo05} confirm that $H(1)\ne H(2) \ne 0.5$ across wide variety of markets and instruments, thus multi-scaling is a characteristic feature of  financial stock markets.

\begin{figure}
   \centering
   \includegraphics[width=\textwidth]{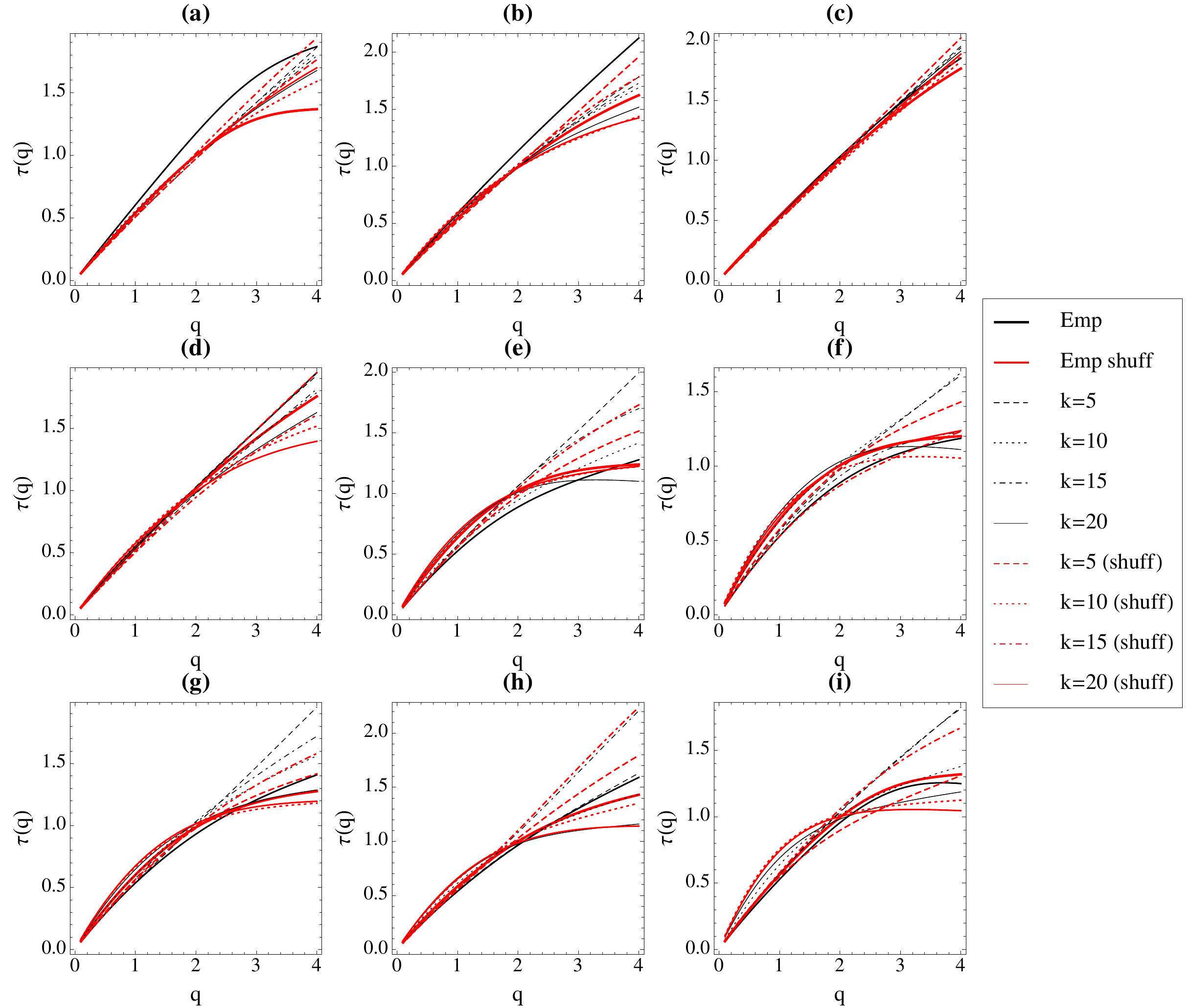}
   \caption{(Color online) Scaling functions $\tau(q)=qH(q)$ as function of $q$ for empirical, MSM simulated data for $k=5,10,15,20$ for $X(t)=p_t$. (a) Dow (b) Nik(c) DM/US (d) US/UK (e) TB1 (f) TB2 (g) TB3 (h) TB5 (i) TB10.}
   \label{plotHq1}
\end{figure}

\begin{figure}
   \centering
   \includegraphics[width=\textwidth]{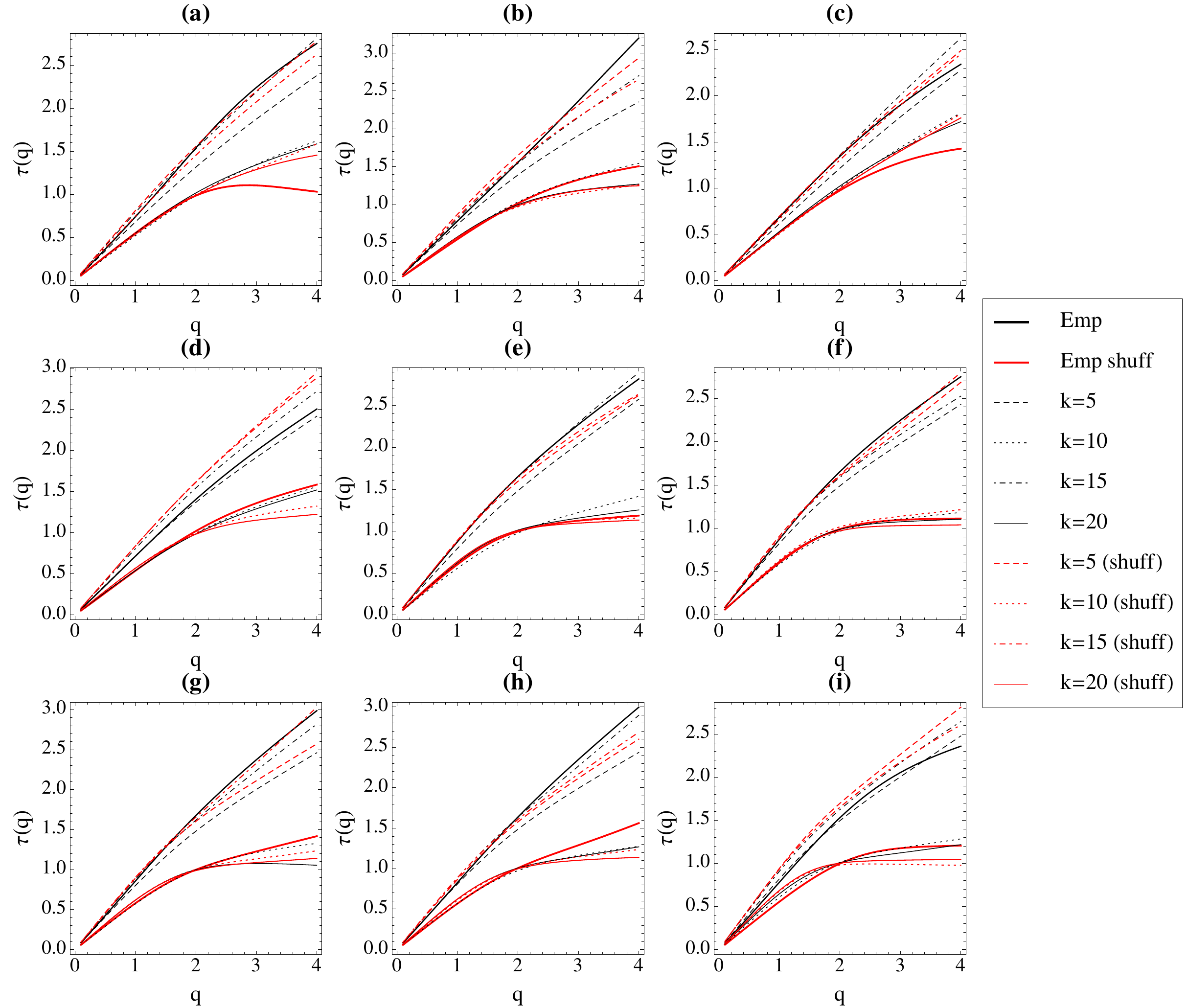}
   \caption{(Color online) Scaling functions $\tau(q)=qH(q)$ as function of $q$ for empirical, MSM simulated data for $k=5,10,15,20$ for $X(t)=\sum |r_{t'}|$. (a) Dow (b) Nik ((c) DM/US (d) US/UK (e) TB1 (f) TB2 (g) TB3 (h) TB5 (i) TB10.}
   \label{plotHq2}
\end{figure}

\begin{figure}
   \centering
   \includegraphics[width=\textwidth]{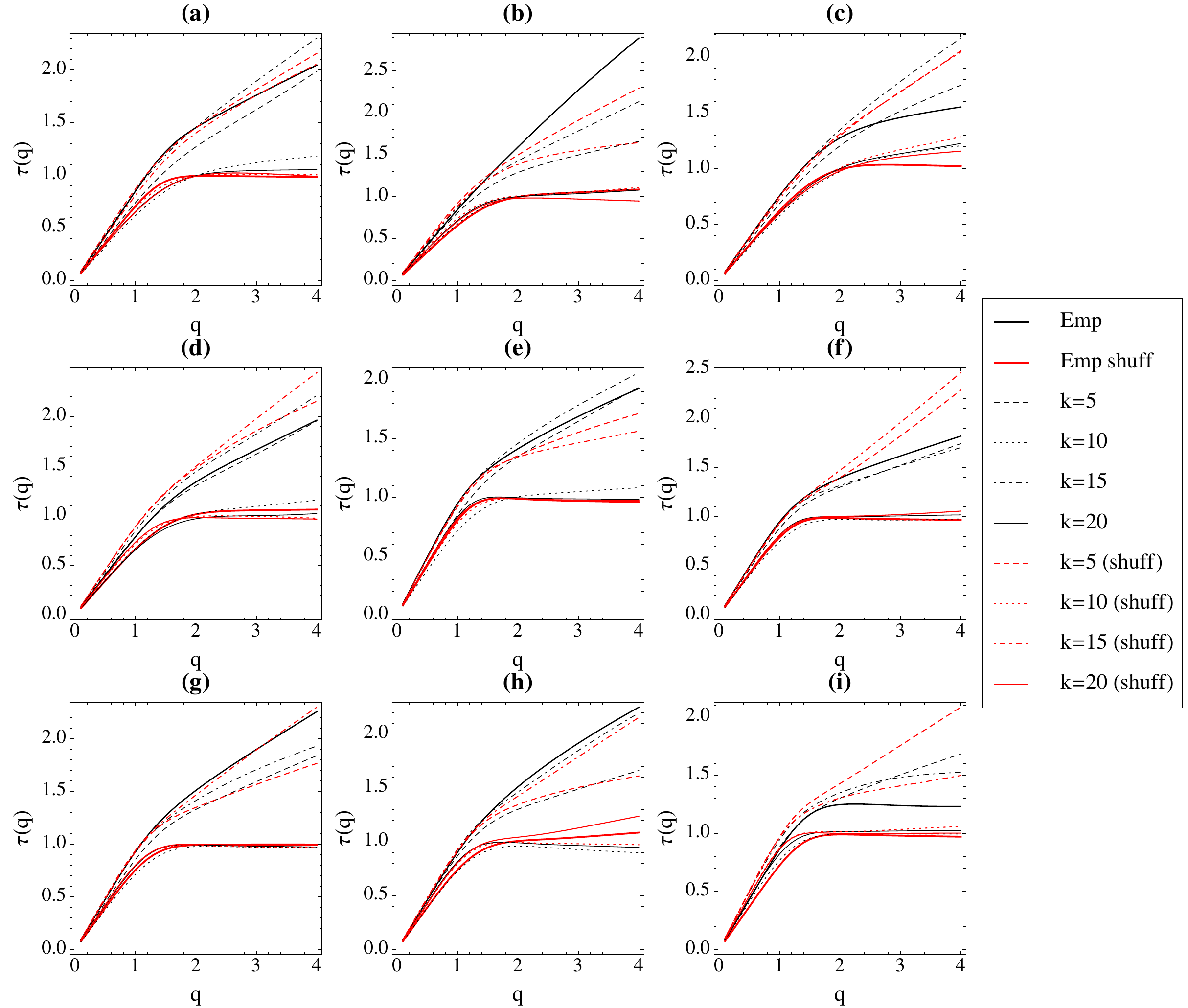}
   \caption{(Color online) Scaling functions $\tau(q)=qH(q)$ as function of $q$ for empirical, MSM simulated data for $k=5,10,15,20$ for $X(t)=\sum r_{t'}^2$. (a) Dow (b) Nik (c) DM/US (d) US/UK (e) TB1 (f) TB2 (g) TB3 (h) TB5 (i) TB10.}
   \label{plotHq3}
\end{figure}

Figures \ref{plotHq1}, \ref{plotHq2} and \ref{plotHq3} show the scaling functions $\tau(q)=qH(q)$ for both empirical and MSM simulated series for prices, absolute returns and squared returns respectively.
Their deviations from linearity  confirm the multi-scaling behavior of all data.
This is a clear signature of deviations from pure Brownian motion and other additive or uni-scaling models.

\subsection{Scaling exponents for prices and volatility}

Once verified that the scaling behavior given in Eq. \ref{K_qtau} holds, we can now discuss the associated average values of generalized Hurst exponent $H(q)$. Our approach is to compare the averages of $H(q)$  of empirical time series with simulations based on the estimated parameters of the MSM models.
The analysis is motivated by our previous results in \cite{LiuAste08,Liu07,Liu08}, where we found some agreement between MSM simulations and empirical data.
We focus on the generalized Hurst exponent (GHE) for $q=\{1,2,3\}$ as for $q>3$ scaling does not hold any longer.
We compute $H(q)$ for the empirical data and compare them to 1000 simulated series for each set of estimated parameters for different values of $k$.

Tables \ref{prices}, \ref{abs} and \ref{sqr} show results for the different stochastic variables, $p_t$, $\sum |r_{t'}|$ and $\sum r_{t'}^2$. 
These tables report $H(q)$ of empirical series and averages of $H(q)$ from 1000 MSM series simulated using the parameter estimates from the Table \ref{gmm}.
If the MSM model describes the scaling behavior of the empirical data well, the identity of $H(q)$ averages from empirical and simulated data should not be statistically rejectable.
These cases are presented in bold. 

Let us start by focusing on the first half of the results titled ``original'' in these tables.
The empirical scaling exponents are varying with different type of data, they are all different among each other as well as different from 0.5, $H(1)\ne H(2)\ne H(3) \ne 0.5$.
We observe that the exponents $H(1)$ and $H(3)$ of simulated series with different $k$ are in most cases slightly different from 0.5, while $H(2)$ can not be statistically distinguished from 0.5.
The fact that $H(2)$ does not change with $k$ and is alway equal to 0.5 might be explained by the fact that MSM is reminiscent of the scaling property for the moments of absolute value of return (price change) rather than price itself.
Therefore, we have performed the same study with stochastic variables constructed as absolute returns and squared returns.
Still, as we can see from the Table \ref{prices}, MSM model does not capture the multi-scaling property of the data in most cases.

Let us now discuss the results for the exponents associated with the absolute returns  $\sum |r_{t'}|$.
We can see in Table \ref{abs} that all estimated scaling exponents, $H(1)$, $H(2)$ and $H(3)$ are in this case greater than 0.5 for both empirical and simulated series.
In particular, a jump in the values of the exponents is observed between $k=5$ and $k>5$ for all series.
We can see that multi-scaling properties of absolute returns are captured well by the MSM model which can not be statistically distinguished from the empirical data for most of the cases.

Let us finally discuss the results for the exponents associated with the squared returns  $\sum r_{t'}^2$ (see Table \ref{sqr}) which are very similar to results from the absolute values case.
In this case, all the values of scaling exponents are much higher than 0.5 and scaling exponents of simulated data correspond well to the empirical data as they can not be statistically distinguished.

In summary these results tell us that the MSM model is able to capture the multi-scaling properties of the volatility, but it is not able to capture the properties of the stock market returns.

\begin{landscape}
\begin{table}[h]
\tiny
\caption{Stochastic variable $p_t$. ``\textit{Emp.}" refers to the empirical exponent values, $k=5$, $k=10$, $k=15$ and $k=20$ refer to the mean and standard deviation of the exponent values based on the 1000 simulated time series. Table contains results for \textbf{original} as well \textbf{shuffled} series. Those cases in which we can not reject identity of the Hurst coefficients obtained for empirical and simulated data on 95\% significance level are in bold. }
\centering
\ra{1.2}
\begin{tabular}{lrrrrrrrrrrrrrrrrr}
\toprule
\multicolumn{18}{c}{\bf{O R I G I N A L}} \\
\hline
&  \multicolumn{5}{c}{$H(1)$} & &  \multicolumn{5}{c}{$H(2)$} & &  \multicolumn{5}{c}{$H(3)$}  \\
\cmidrule{2-6} \cmidrule{8-12} \cmidrule{14-18}
&Emp. & $k=5$ & $k=10$ & $k=15$ & $k=20$ &  &Emp. & $k=5$ & $k=10$ & $k=15$ & $k=20$ &  &Emp. & $k=5$ & $k=10$ & $k=15$ & $k=20$  \\
\hline
Dow &0.602 & 0.515 & 0.513 & 0.513 & 0.513 &    &0.586 & 0.500 & 0.500 & 0.499 & 0.499 &    &0.543 & 0.485 & 0.487 & 0.486 & 0.485 \\
  &(0.038) & (0.010) & (0.010) & (0.010) & (0.010) &    &(0.031) & (0.010) & (0.012) & (0.013) & (0.014) &    &(0.026) & (0.013) & (0.018) & (0.020) & (0.021) \\
Nik  &0.587 & 0.524 & 0.524 & 0.524 & 0.524 &    &0.566 & 0.499 & 0.499 & 0.498 & 0.497 &    &0.547 & 0.477 & 0.474 & 0.473 & 0.471 \\
  &(0.019) & (0.010) & (0.011) & (0.011) & (0.011) &    &(0.024) & (0.011) & (0.016) & (0.018) & (0.018) &    &(0.033) & (0.015) & (0.027) & (0.031) & (0.030) \\
UK &0.539 & 0.507 & 0.505 & 0.505 & 0.505 &    &0.516 & 0.499 & 0.499 & 0.499 & 0.499 &    &0.492 & \textbf{0.492} & \textbf{0.494} & \textbf{0.494} & 0.494 \\
  &(0.003) & (0.009) & (0.010) & (0.010) & (0.009) &    &(0.005) & (0.009) & (0.010) & (0.010) & (0.010) &    &(0.004) & (0.011) & (0.012) & (0.013) & (0.013) \\
Aud &0.534 & \textbf{0.521} & 0.516 & 0.515 & 0.515 &    &0.511 & \textbf{0.499} & \textbf{0.499} & \textbf{0.498} & \textbf{0.498} &    &0.495 & \textbf{0.479} & \textbf{0.482} & \textbf{0.482} & 0.482 \\
  &(0.005) & (0.010) & (0.010) & (0.010) & (0.011) &    &(0.004) & (0.011) & (0.014) & (0.015) & (0.015) &    &(0.006) & (0.014) & (0.021) & (0.024) & (0.024) \\
TB1 &0.525 & 0.553 & 0.549 & 0.550 & 0.550 &    &0.444 & 0.500 & 0.497 & 0.498 & 0.498 &    &0.371 & 0.459 & 0.451 & \textbf{0.453} & \textbf{0.451} \\
  &(0.005) & (0.011) & (0.014) & (0.016) & (0.015) &    &(0.016) & (0.015) & (0.026) & (0.031) & (0.031) &    &(0.032) & (0.022) & (0.045) & (0.052) & (0.052) \\
TB2 &0.528 & 0.569 & 0.566 & 0.566 & 0.566 &    &0.443 & 0.498 & 0.497 & 0.495 & 0.495 &    &0.362 & 0.448 & 0.441 & \textbf{0.437} & \textbf{0.437} \\
  &(0.003) & (0.012) & (0.015) & (0.018) & (0.018) &    &(0.008) & (0.016) & (0.032) & (0.038) & (0.039) &    &(0.013) & (0.024) & (0.055) & (0.062) & (0.062) \\
TB3 &0.536 & \textbf{0.550} & \textbf{0.548} & \textbf{0.548} & \textbf{0.547} &    &0.465 & 0.499 & \textbf{0.497} & \textbf{0.495} & \textbf{0.495} &    &0.402 & 0.459 & \textbf{0.453} & \textbf{0.450} & \textbf{0.449} \\
  &(0.004) & (0.010) & (0.014) & (0.014) & (0.015) &    &(0.009) & (0.013) & (0.025) & (0.028) & (0.030) &    &(0.014) & (0.020) & (0.042) & (0.050) & (0.051) \\
TB5 &0.537 & 0.558 & 0.554 & \textbf{0.555} & \textbf{0.554} &    &0.481 & \textbf{0.499} & \textbf{0.496} & \textbf{0.495} & \textbf{0.496} &    &0.434 & \textbf{0.455} & \textbf{0.445} & \textbf{0.444} & \textbf{0.445} \\
  &(0.003) & (0.011) & (0.014) & (0.016) & (0.015) &    &(0.005) & (0.015) & (0.027) & (0.032) & (0.033) &    &(0.010) & (0.022) & (0.048) & (0.055) & (0.057) \\
TB10 &0.526 & 0.565 & 0.564 & 0.565 & 0.564 &    &0.477 & 0.499 & \textbf{0.496} & \textbf{0.495} & \textbf{0.494} &    &0.404 & 0.451 & \textbf{0.441} & \textbf{0.436} & \textbf{0.437} \\
  &(0.003) & (0.012) & (0.016) & (0.017) & (0.018) &    &(0.003) & (0.015) & (0.031) & (0.036) & (0.038) &    &(0.007) & (0.022) & (0.052) & (0.061) & (0.063) \\
\hline
\multicolumn{18}{c}{\bf{S H U F F L E D}} \\
\hline
&  \multicolumn{5}{c}{$H(1)$} & &  \multicolumn{5}{c}{$H(2)$} & &  \multicolumn{5}{c}{$H(3)$}  \\
\cmidrule{2-6} \cmidrule{8-12} \cmidrule{14-18}
&Emp. & $k=5$ & $k=10$ & $k=15$ & $k=20$ &  &Emp. & $k=5$ & $k=10$ & $k=15$ & $k=20$ &  &Emp. & $k=5$ & $k=10$ & $k=15$ & $k=20$  \\
\hline
Dow &0.543 & \textbf{0.531} & \textbf{0.544} & 0.555 & \textbf{0.541} &    &0.502 & \textbf{0.503} & \textbf{0.502} & \textbf{0.501} & \textbf{0.500} &    &0.419 & 0.478 & 0.455 & 0.455 & 0.463 \\
  &(0.003) & (0.009) & (0.013) & (0.004) & (0.007) &    &(0.005) & (0.004) & (0.013) & (0.006) & (0.008) &    &(0.008) & (0.004) & (0.012) & (0.012) & (0.009) \\
Nik  &0.541 & \textbf{0.546} & 0.588 & 0.605 & 0.580 &    &0.499 & \textbf{0.499} & \textbf{0.497} & \textbf{0.494} & \textbf{0.501} &    &0.455 & \textbf{0.458} & 0.419 & 0.403 & 0.436 \\
  &(0.010) & (0.005) & (0.004) & (0.004) & (0.006) &    &(0.009) & (0.005) & (0.003) & (0.006) & (0.005) &    &(0.009) & (0.006) & (0.005) & (0.012) & (0.004) \\
UK &0.524 & \textbf{0.519} & 0.515 & \textbf{0.523} & \textbf{0.521} &    &0.492 & \textbf{0.503} & \textbf{0.499} & 0.508 & 0.502 &    &0.458 & 0.486 & 0.484 & 0.493 & \textbf{0.483} \\
  &(0.004) & (0.008) & (0.004) & (0.009) & (0.002) &    &(0.003) & (0.011) & (0.004) & (0.008) & (0.006) &    &(0.004) & (0.014) & (0.008) & (0.008) & (0.011) \\
Aud &0.547 & \textbf{0.538} & \textbf{0.553} & \textbf{0.547} & \textbf{0.553} &    &0.506 & \textbf{0.500} & \textbf{0.501} & \textbf{0.498} & \textbf{0.493} &    &0.464 & \textbf{0.467} & \textbf{0.455} & 0.451 & 0.438 \\
  &(0.014) & (0.009) & (0.010) & (0.004) & (0.004) &    &(0.011) & (0.013) & (0.008) & (0.002) & (0.005) &    &(0.008) & (0.017) & (0.007) & (0.003) & (0.005) \\
TB1 &0.636 & 0.601 & \textbf{0.653} & 0.656 & 0.662 &    &0.507 & \textbf{0.501} & 0.498 & 0.494 & 0.497 &    &0.390 & 0.419 & \textbf{0.376} & \textbf{0.375} & 0.368 \\
  &(0.011) & (0.004) & (0.013) & (0.008) & (0.004) &    &(0.002) & (0.006) & (0.006) & (0.008) & (0.002) &    &(0.016) & (0.008) & (0.004) & (0.011) & (0.012) \\
TB2 &0.623 & \textbf{0.629} & 0.679 & 0.722 & 0.670 &    &0.500 & \textbf{0.499} & \textbf{0.502} & \textbf{0.497} & \textbf{0.498} &    &0.393 & \textbf{0.407} & \textbf{0.377} & 0.351 & 0.379 \\
  &(0.005) & (0.005) & (0.010) & (0.000) & (0.004) &    &(0.004) & (0.006) & (0.018) & (0.003) & (0.004) &    &(0.009) & (0.008) & (0.022) & (0.007) & (0.011) \\
TB3 &0.607 & 0.591 & 0.679 & 0.715 & 0.651 &    &0.501 & \textbf{0.497} & \textbf{0.497} & \textbf{0.494} & \textbf{0.494} &    &0.408 & 0.423 & 0.359 & 0.355 & 0.378 \\
  &(0.010) & (0.005) & (0.002) & (0.009) & (0.004) &    &(0.005) & (0.002) & (0.003) & (0.009) & (0.007) &    &(0.004) & (0.001) & (0.009) & (0.012) & (0.011) \\
TB5 &0.588 & 0.614 & 0.673 & 0.689 & 0.670 &    &0.503 & \textbf{0.502} & \textbf{0.502} & \textbf{0.503} & \textbf{0.505} &    &0.426 & \textbf{0.416} & 0.374 & 0.370 & 0.395 \\
  &(0.001) & (0.007) & (0.011) & (0.006) & (0.009) &    &(0.006) & (0.007) & (0.007) & (0.002) & (0.012) &    &(0.013) & (0.010) & (0.007) & (0.004) & (0.013) \\
TB10 &0.571 & 0.629 & 0.669 & 0.743 & 0.712 &    &0.508 & \textbf{0.503} & \textbf{0.509} & 0.491 & \textbf{0.502} &    &0.420 & \textbf{0.413} & 0.396 & 0.337 & 0.360 \\
  &(0.008) & (0.003) & (0.002) & (0.012) & (0.006) &    &(0.006) & (0.001) & (0.004) & (0.005) & (0.017) &    &(0.005) & (0.003) & (0.009) & (0.002) & (0.032) \\
 \bottomrule
\end{tabular}
\label{prices}
\end{table}
\end{landscape}

\begin{landscape}
\begin{table}[h]
\tiny
\caption{Stochastic variable $\sum |r_{t'}|$. ``\textit{Emp.}" refers to the empirical exponent values, $k=5$, $k=10$, $k=15$ and $k=20$ refer to the mean and standard deviation of the exponent values based on the 1000 simulated time series. Table contains results for \textbf{original} as well \textbf{shuffled} series. Those cases in which we can not reject identity of the Hurst coefficients obtained for empirical and simulated data on 95\% significance level are in bold.}
\centering
\ra{1.2}
\begin{tabular}{lrrrrrrrrrrrrrrrrr}
\toprule
\multicolumn{18}{c}{\bf{O R I G I N A L}} \\
\hline
&  \multicolumn{5}{c}{$H(1)$} & &  \multicolumn{5}{c}{$H(2)$} & &  \multicolumn{5}{c}{$H(3)$}  \\
\cmidrule{2-6} \cmidrule{8-12} \cmidrule{14-18}
&Emp. & $k=5$ & $k=10$ & $k=15$ & $k=20$ &  &Emp. & $k=5$ & $k=10$ & $k=15$ & $k=20$ &  &Emp. & $k=5$ & $k=10$ & $k=15$ & $k=20$  \\
\hline
Dow &0.738 & 0.682 & \textbf{0.761} & 0.787 & 0.789 &    &0.769 & 0.658 & 0.735 & \textbf{0.758} & \textbf{0.759} &    &0.747 & 0.619 & 0.695 & \textbf{0.715} & 0.716 \\
  &(0.018) & (0.010) & (0.017) & (0.027) & (0.026) &    &(0.016) & (0.009) & (0.016) & (0.023) & (0.023) &    &(0.009) & (0.013) & (0.019) & (0.025) & (0.025) \\
Nik  &0.781 & 0.726 & 0.829 & 0.848 & 0.850 &    &0.778 & 0.690 & \textbf{0.784} & \textbf{0.800} & \textbf{0.802} &    &0.789 & 0.640 & 0.726 & 0.740 & \textbf{0.742} \\
  &(0.024) & (0.010) & (0.016) & (0.024) & (0.024) &    &(0.019) & (0.009) & (0.015) & (0.021) & (0.020) &    &(0.018) & (0.014) & (0.021) & (0.027) & (0.024) \\
UK &0.686 & 0.618 & \textbf{0.669} & \textbf{0.689} & \textbf{0.693} &    &0.672 & 0.606 & \textbf{0.657} & \textbf{0.676} & \textbf{0.680} &    &0.633 & 0.581 & \textbf{0.634} & \textbf{0.651} & \textbf{0.655} \\
  &(0.031) & (0.010) & (0.016) & (0.026) & (0.027) &    &(0.029) & (0.009) & (0.015) & (0.024) & (0.025) &    &(0.025) & (0.012) & (0.017) & (0.023) & (0.025) \\
Aud &0.713 & \textbf{0.711} & 0.788 & 0.811 & 0.813 &    &0.702 & \textbf{0.680} & 0.756 & 0.776 & 0.778 &    &0.665 & \textbf{0.634} & \textbf{0.710} & 0.727 & \textbf{0.730} \\
  &(0.026) & (0.010) & (0.016) & (0.025) & (0.025) &    &(0.031) & (0.009) & (0.015) & (0.022) & (0.022) &    &(0.036) & (0.013) & (0.019) & (0.025) & (0.024) \\
TB1 &0.872 & 0.798 & \textbf{0.883} & \textbf{0.900} & \textbf{0.902} &    &0.825 & 0.734 & \textbf{0.811} & \textbf{0.823} & \textbf{0.824} &    &0.757 & 0.660 & \textbf{0.733} & \textbf{0.743} & 0.744 \\
  &(0.022) & (0.008) & (0.013) & (0.019) & (0.019) &    &(0.025) & (0.009) & (0.014) & (0.018) & (0.017) &    &(0.034) & (0.015) & (0.028) & (0.031) & (0.030) \\
TB2 &0.874 & 0.821 & \textbf{0.902} & 0.916 & 0.916 &    &0.826 & 0.743 & \textbf{0.812} & \textbf{0.823} & \textbf{0.823} &    &0.749 & 0.660 & \textbf{0.725} & \textbf{0.736} & 0.735 \\
  &(0.020) & (0.008) & (0.012) & (0.016) & (0.017) &    &(0.023) & (0.009) & (0.015) & (0.017) & (0.017) &    &(0.031) & (0.017) & (0.032) & (0.034) & (0.035) \\
TB3 &0.855 & 0.793 & \textbf{0.881} & 0.898 & 0.899 &    &0.838 & 0.731 & \textbf{0.810} & \textbf{0.823} & \textbf{0.824} &    &0.791 & 0.659 & \textbf{0.734} & \textbf{0.745} & 0.745 \\
  &(0.022) & (0.008) & (0.014) & (0.018) & (0.019) &    &(0.025) & (0.009) & (0.014) & (0.016) & (0.017) &    &(0.031) & (0.015) & (0.028) & (0.029) & (0.029) \\
TB5 &0.826 & \textbf{0.806} & 0.889 & 0.905 & 0.906 &    &0.818 & 0.737 & \textbf{0.812} & \textbf{0.823} & \textbf{0.824} &    &0.782 & 0.660 & \textbf{0.732} & \textbf{0.740} & 0.742 \\
  &(0.024) & (0.009) & (0.013) & (0.017) & (0.018) &    &(0.027) & (0.009) & (0.014) & (0.017) & (0.016) &    &(0.032) & (0.015) & (0.029) & (0.032) & (0.031) \\
TB10 &0.781 & \textbf{0.816} & 0.900 & 0.915 & 0.916 &    &0.764 & \textbf{0.741} & 0.814 & 0.822 & 0.824 &    &0.684 & \textbf{0.660} & \textbf{0.729} & \textbf{0.733} & \textbf{0.736} \\
  &(0.028) & (0.008) & (0.012) & (0.017) & (0.017) &    &(0.033) & (0.009) & (0.014) & (0.017) & (0.016) &    &(0.049) & (0.017) & (0.030) & (0.037) & (0.033) \\
\hline
\multicolumn{18}{c}{\bf{S H U F F L E D}} \\
\hline
&  \multicolumn{5}{c}{$H(1)$} & &  \multicolumn{5}{c}{$H(2)$} & &  \multicolumn{5}{c}{$H(3)$}  \\
\cmidrule{2-6} \cmidrule{8-12} \cmidrule{14-18}
&Emp. & $k=5$ & $k=10$ & $k=15$ & $k=20$ &  &Emp. & $k=5$ & $k=10$ & $k=15$ & $k=20$ &  &Emp. & $k=5$ & $k=10$ & $k=15$ & $k=20$  \\
\hline
Dow &0.548 & 0.522 & \textbf{0.540} & 0.537 & 0.531 &    &0.495 & \textbf{0.496} & \textbf{0.495} & \textbf{0.492} & \textbf{0.496} &    &0.377 & 0.452 & 0.418 & 0.424 & 0.440 \\
  &(0.006) & (0.009) & (0.003) & (0.005) & (0.009) &    &(0.008) & (0.009) & (0.003) & (0.008) & (0.006) &    &(0.007) & (0.009) & (0.009) & (0.012) & (0.002) \\
Nik  &0.541 & \textbf{0.540} & 0.573 & 0.576 & 0.558 &    &0.500 & \textbf{0.501} & \textbf{0.507} & \textbf{0.494} & \textbf{0.501} &    &0.429 & \textbf{0.444} & \textbf{0.414} & 0.388 & 0.424 \\
  &(0.010) & (0.010) & (0.001) & (0.003) & (0.002) &    &(0.010) & (0.010) & (0.002) & (0.002) & (0.005) &    &(0.014) & (0.010) & (0.004) & (0.009) & (0.014) \\
UK &0.530 & \textbf{0.517} & 0.515 & 0.507 & \textbf{0.521} &    &0.502 & \textbf{0.501} & \textbf{0.500} & \textbf{0.492} & \textbf{0.500} &    &0.445 & 0.471 & 0.467 & 0.463 & 0.461 \\
  &(0.007) & (0.016) & (0.007) & (0.007) & (0.012) &    &(0.007) & (0.016) & (0.005) & (0.006) & (0.015) &    &(0.007) & (0.017) & (0.003) & (0.004) & (0.020) \\
Aud &0.532 & \textbf{0.528} & \textbf{0.541} & \textbf{0.548} & \textbf{0.540} &    &0.498 & \textbf{0.496} & \textbf{0.498} & \textbf{0.505} & 0.488 &    &0.437 & \textbf{0.444} & \textbf{0.433} & \textbf{0.439} & 0.409 \\
  &(0.012) & (0.006) & (0.003) & (0.008) & (0.004) &    &(0.009) & (0.007) & (0.007) & (0.006) & (0.000) &    &(0.007) & (0.008) & (0.011) & (0.004) & (0.005) \\
TB1 &0.615 & 0.579 & \textbf{0.622} & \textbf{0.625} & 0.639 &    &0.497 & \textbf{0.505} & \textbf{0.502} & \textbf{0.496} & \textbf{0.495} &    &0.357 & 0.408 & \textbf{0.370} & \textbf{0.359} & 0.346 \\
  &(0.008) & (0.008) & (0.004) & (0.008) & (0.002) &    &(0.012) & (0.010) & (0.005) & (0.016) & (0.006) &    &(0.014) & (0.011) & (0.007) & (0.026) & (0.013) \\
TB2 &0.617 & 0.600 & \textbf{0.625} & 0.669 & \textbf{0.622} &    &0.504 & \textbf{0.499} & \textbf{0.495} & \textbf{0.492} & \textbf{0.509} &    &0.376 & \textbf{0.390} & \textbf{0.355} & 0.333 & 0.390 \\
  &(0.007) & (0.001) & (0.010) & (0.011) & (0.011) &    &(0.011) & (0.002) & (0.008) & (0.004) & (0.009) &    &(0.021) & (0.002) & (0.007) & (0.004) & (0.013) \\
TB3 &0.585 & \textbf{0.572} & 0.645 & 0.662 & 0.619 &    &0.492 & \textbf{0.501} & 0.507 & \textbf{0.501} & \textbf{0.502} &    &0.376 & 0.412 & \textbf{0.367} & \textbf{0.353} & 0.374 \\
  &(0.002) & (0.015) & (0.004) & (0.003) & (0.010) &    &(0.009) & (0.016) & (0.003) & (0.005) & (0.011) &    &(0.021) & (0.014) & (0.008) & (0.011) & (0.020) \\
TB5 &0.577 & \textbf{0.579} & 0.632 & 0.647 & 0.623 &    &0.501 & \textbf{0.490} & \textbf{0.501} & \textbf{0.498} & \textbf{0.499} &    &0.405 & \textbf{0.387} & 0.360 & 0.356 & 0.375 \\
  &(0.007) & (0.007) & (0.005) & (0.005) & (0.008) &    &(0.008) & (0.008) & (0.009) & (0.004) & (0.009) &    &(0.009) & (0.013) & (0.012) & (0.011) & (0.019) \\
TB10 &0.557 & 0.588 & 0.622 & 0.702 & 0.657 &    &0.496 & \textbf{0.502} & \textbf{0.496} & \textbf{0.492} & \textbf{0.500} &    &0.379 & \textbf{0.404} & \textbf{0.360} & 0.329 & 0.344 \\
  &(0.005) & (0.002) & (0.010) & (0.009) & (0.003) &    &(0.007) & (0.006) & (0.004) & (0.005) & (0.006) &    &(0.014) & (0.016) & (0.011) & (0.003) & (0.015) \\
   \bottomrule
\end{tabular}
\label{abs}
\end{table}
\end{landscape}

\begin{landscape}
\begin{table}[h]
\tiny
\caption{Stochastic variable $\sum r_{t'}^2$. ``\textit{Emp.}" refers to the empirical exponent values, $k=5$, $k=10$, $k=15$ and $k=20$ refer to the mean and standard deviation of the exponent values based on the 1000 simulated time series. Table contains results for \textbf{original} as well \textbf{shuffled} series. Those cases in which we can not reject identity of the Hurst coefficients obtained for empirical and simulated data on 95\% significance level are in bold.}
\centering
\ra{1.2}
\begin{tabular}{lrrrrrrrrrrrrrrrrr}
\toprule
\multicolumn{18}{c}{\bf{O R I G I N A L}} \\
\hline
&  \multicolumn{5}{c}{$H(1)$} & &  \multicolumn{5}{c}{$H(2)$} & &  \multicolumn{5}{c}{$H(3)$}  \\
\cmidrule{2-6} \cmidrule{8-12} \cmidrule{14-18}
&Emp. & $k=5$ & $k=10$ & $k=15$ & $k=20$ &  &Emp. & $k=5$ & $k=10$ & $k=15$ & $k=20$ &  &Emp. & $k=5$ & $k=10$ & $k=15$ & $k=20$  \\
\hline
Dow &0.847 & 0.750 & 0.821 & \textbf{0.843} & \textbf{0.845} &    &0.725 & 0.633 & 0.696 & \textbf{0.711} & \textbf{0.711} &    &0.586 & 0.514 & \textbf{0.579} & \textbf{0.592} & 0.591 \\
  &(0.010) & (0.009) & (0.016) & (0.023) & (0.023) &    &(0.008) & (0.012) & (0.018) & (0.022) & (0.022) &    &(0.013) & (0.026) & (0.040) & (0.044) & (0.043) \\
Nik  &0.840 & 0.791 & 0.879 & 0.895 & 0.896 &    &0.794 & 0.647 & 0.709 & 0.718 & 0.718 &    &0.756 & 0.518 & 0.577 & 0.585 & 0.586 \\
  &(0.012) & (0.009) & (0.015) & (0.020) & (0.019) &    &(0.015) & (0.013) & (0.023) & (0.027) & (0.027) &    &(0.022) & (0.029) & (0.049) & (0.054) & (0.054) \\
UK &0.755 & 0.687 & \textbf{0.735} & \textbf{0.753} & \textbf{0.758} &    &0.637 & 0.599 & \textbf{0.648} & \textbf{0.664} & \textbf{0.667} &    &0.486 & \textbf{0.498} & 0.552 & 0.565 & 0.568 \\
  &(0.017) & (0.009) & (0.015) & (0.025) & (0.025) &    &(0.018) & (0.010) & (0.016) & (0.021) & (0.022) &    &(0.015) & (0.021) & (0.028) & (0.030) & (0.033) \\
Aud &0.786 & \textbf{0.778} & 0.845 & 0.864 & 0.866 &    &0.670 & \textbf{0.644} & \textbf{0.705} & \textbf{0.716} & \textbf{0.719} &    &0.555 & \textbf{0.517} & \textbf{0.582} & \textbf{0.590} & \textbf{0.597} \\
  &(0.014) & (0.009) & (0.015) & (0.022) & (0.022) &    &(0.029) & (0.012) & (0.019) & (0.023) & (0.022) &    &(0.038) & (0.027) & (0.041) & (0.047) & (0.045) \\
TB1 &0.945 & 0.857 & 0.926 & \textbf{0.938} & \textbf{0.939} &    &0.706 & 0.653 & \textbf{0.696} & \textbf{0.701} & \textbf{0.701} &    &0.562 & \textbf{0.508} & \textbf{0.554} & \textbf{0.560} & 0.560 \\
  &(0.007) & (0.007) & (0.012) & (0.015) & (0.015) &    &(0.026) & (0.016) & (0.036) & (0.039) & (0.039) &    &(0.030) & (0.032) & (0.065) & (0.067) & (0.069) \\
TB2 &0.945 & 0.877 & \textbf{0.941} & \textbf{0.951} & \textbf{0.951} &    &0.695 & 0.649 & \textbf{0.683} & \textbf{0.691} & \textbf{0.689} &    &0.539 & \textbf{0.502} & \textbf{0.538} & \textbf{0.550} & 0.547 \\
  &(0.006) & (0.007) & (0.010) & (0.013) & (0.013) &    &(0.028) & (0.018) & (0.040) & (0.044) & (0.044) &    &(0.043) & (0.035) & (0.067) & (0.072) & (0.073) \\
TB3 &0.927 & 0.852 & \textbf{0.924} & \textbf{0.936} & \textbf{0.937} &    &0.754 & 0.653 & \textbf{0.697} & \textbf{0.704} & \textbf{0.703} &    &0.632 & 0.509 & \textbf{0.556} & \textbf{0.565} & 0.563 \\
  &(0.008) & (0.008) & (0.012) & (0.014) & (0.015) &    &(0.021) & (0.016) & (0.036) & (0.036) & (0.037) &    &(0.020) & (0.033) & (0.066) & (0.065) & (0.065) \\
TB5 &0.901 & 0.864 & 0.931 & 0.942 & 0.943 &    &0.755 & 0.652 & \textbf{0.693} & \textbf{0.697} & \textbf{0.698} &    &0.638 & 0.507 & \textbf{0.552} & \textbf{0.556} & 0.558 \\
  &(0.011) & (0.007) & (0.012) & (0.015) & (0.014) &    &(0.033) & (0.016) & (0.037) & (0.042) & (0.040) &    &(0.052) & (0.033) & (0.066) & (0.071) & (0.066) \\
TB10 &0.860 & \textbf{0.873} & 0.940 & 0.950 & 0.951 &    &0.623 & \textbf{0.650} & \textbf{0.688} & \textbf{0.687} & \textbf{0.690} &    &0.413 & 0.503 & 0.547 & 0.545 & \textbf{0.547} \\
  &(0.015) & (0.007) & (0.011) & (0.013) & (0.013) &    &(0.031) & (0.017) & (0.038) & (0.046) & (0.042) &    &(0.032) & (0.033) & (0.066) & (0.075) & (0.069) \\
  \hline
\multicolumn{18}{c}{\bf{S H U F F L E D}} \\
\hline
&  \multicolumn{5}{c}{$H(1)$} & &  \multicolumn{5}{c}{$H(2)$} & &  \multicolumn{5}{c}{$H(3)$}  \\
\cmidrule{2-6} \cmidrule{8-12} \cmidrule{14-18}
&Emp. & $k=5$ & $k=10$ & $k=15$ & $k=20$ &  &Emp. & $k=5$ & $k=10$ & $k=15$ & $k=20$ &  &Emp. & $k=5$ & $k=10$ & $k=15$ & $k=20$  \\
\hline
Dow &0.720 & 0.616 & 0.672 & 0.673 & 0.650 &    &0.500 & \textbf{0.497} & 0.494 & 0.495 & \textbf{0.502} &    &0.333 & 0.365 & \textbf{0.329} & \textbf{0.339} & 0.359 \\
  &(0.003) & (0.007) & (0.003) & (0.001) & (0.007) &    &(0.001) & (0.007) & (0.003) & (0.001) & (0.005) &    &(0.003) & (0.007) & (0.007) & (0.003) & (0.005) \\
Nik  &0.654 & \textbf{0.650} & 0.715 & 0.750 & 0.706 &    &0.492 & \textbf{0.497} & \textbf{0.511} & \textbf{0.501} & \textbf{0.496} &    &0.332 & \textbf{0.360} & \textbf{0.356} & \textbf{0.328} & 0.335 \\
  &(0.006) & (0.004) & (0.003) & (0.005) & (0.006) &    &(0.011) & (0.005) & (0.016) & (0.008) & (0.010) &    &(0.016) & (0.016) & (0.032) & (0.006) & (0.015) \\
UK &0.632 & 0.598 & 0.601 & 0.602 & 0.610 &    &0.498 & \textbf{0.496} & \textbf{0.498} & \textbf{0.503} & \textbf{0.497} &    &0.347 & 0.377 & 0.374 & 0.397 & 0.370 \\
  &(0.003) & (0.008) & (0.007) & (0.009) & (0.007) &    &(0.004) & (0.010) & (0.004) & (0.008) & (0.008) &    &(0.008) & (0.018) & (0.002) & (0.005) & (0.011) \\
Aud &0.658 & 0.644 & \textbf{0.661} & 0.676 & 0.678 &    &0.505 & \textbf{0.500} & \textbf{0.494} & \textbf{0.503} & \textbf{0.492} &    &0.358 & \textbf{0.355} & \textbf{0.347} & \textbf{0.345} & 0.336 \\
  &(0.009) & (0.004) & (0.004) & (0.006) & (0.005) &    &(0.012) & (0.007) & (0.013) & (0.011) & (0.010) &    &(0.026) & (0.013) & (0.026) & (0.016) & (0.013) \\
TB1 &0.805 & 0.716 & \textbf{0.805} & \textbf{0.808} & 0.831 &    &0.499 & \textbf{0.492} & \textbf{0.503} & \textbf{0.495} & \textbf{0.497} &    &0.328 & \textbf{0.330} & \textbf{0.335} & \textbf{0.327} & 0.328 \\
  &(0.008) & (0.004) & (0.005) & (0.003) & (0.002) &    &(0.006) & (0.013) & (0.011) & (0.002) & (0.003) &    &(0.004) & (0.018) & (0.013) & (0.005) & (0.005) \\
TB2 &0.786 & 0.760 & 0.801 & 0.860 & \textbf{0.788} &    &0.494 & 0.505 & \textbf{0.503} & \textbf{0.499} & \textbf{0.500} &    &0.327 & 0.345 & \textbf{0.334} & \textbf{0.331} & 0.330 \\
  &(0.007) & (0.000) & (0.004) & (0.002) & (0.003) &    &(0.003) & (0.005) & (0.007) & (0.002) & (0.004) &    &(0.003) & (0.009) & (0.010) & (0.003) & (0.004) \\
TB3 &0.764 & 0.716 & 0.825 & 0.849 & 0.793 &    &0.499 & \textbf{0.502} & \textbf{0.505} & 0.495 & \textbf{0.499} &    &0.328 & \textbf{0.345} & \textbf{0.341} & \textbf{0.327} & 0.336 \\
  &(0.004) & (0.003) & (0.007) & (0.001) & (0.004) &    &(0.003) & (0.010) & (0.013) & (0.001) & (0.008) &    &(0.009) & (0.018) & (0.019) & (0.001) & (0.013) \\
TB5 &0.736 & \textbf{0.737} & 0.816 & 0.843 & 0.797 &    &0.511 & \textbf{0.495} & \textbf{0.497} & \textbf{0.499} & \textbf{0.504} &    &0.362 & \textbf{0.338} & \textbf{0.329} & \textbf{0.331} & 0.346 \\
  &(0.011) & (0.005) & (0.002) & (0.003) & (0.003) &    &(0.017) & (0.001) & (0.006) & (0.005) & (0.008) &    &(0.028) & (0.009) & (0.008) & (0.007) & (0.026) \\
TB10 &0.721 & 0.747 & 0.808 & 0.893 & 0.849 &    &0.493 & \textbf{0.491} & \textbf{0.500} & \textbf{0.496} & \textbf{0.497} &    &0.322 & \textbf{0.330} & \textbf{0.334} & \textbf{0.326} & 0.330 \\
  &(0.006) & (0.005) & (0.002) & (0.002) & (0.000) &    &(0.013) & (0.001) & (0.005) & (0.002) & (0.001) &    &(0.016) & (0.006) & (0.007) & (0.001) & (0.000) \\
 \bottomrule
\end{tabular}
\label{sqr}
\end{table}
\end{landscape}

\subsection{The source of multifractality}

Let us now proceed with the same analysis as in the previous section, but this time shuffling the data randomly to better understand the source of multifractality. 
One can argue that, if the behavior described in the previous section is caused only by the long-range correlations, then the randomly shuffled series should show unifractality, and $\Delta H_{Shuff}=H_{Shuff}(1)-H_{Shuff}(3)=0$, where $H_{Shuff}(q)$ is the generalized Hurst exponent of the shuffled time series.
If the multifractality is caused only by the broad distribution, than instead we must have $\Delta H=\Delta H_{Shuff}>0$.

We follow our analysis by shuffling the time series 33 times\footnote{Note that we have tried different numbers of shuffled time series from 2 to 100, but results do not change. These results are available upon request from authors.} to achieve robustness of the results. In the shuffling procedure the values of the time series are put into random order destroying all correlations. We shuffle empirical as well as simulated series, we check the scaling behavior and we report the results for the shuffled data in Figures \ref{plotK}, \ref{plotKabs}, \ref{plotKsqr}, \ref{plotMSM1}, \ref{plotMSM2}, \ref{plotMSM3}. Scaling of shuffled series is provided by the red lines (color online) and we can confirm that scaling holds for all considered $q$ values thus we can proceed in computing the scaling exponents\footnote{Similar scaling behaviors have been also found for the other simulated time series based on all pairs on estimated coefficients from the Table \ref{gmm} and for different considered stochastic variables.}.

If we look at Tables \ref{prices}, \ref{abs} and \ref{sqr} where we report the results for the prices, absolute returns and squared returns respectively, we can compare the scaling behavior of ``original" series with the ``shuffled" ones.
We note that the empirical scaling exponents are varying with different type of data even after shuffling, although they are generally closer to 0.5.

Let us start discussing the results for the prices.
After shuffling, empirical data show even larger degree of  multifractality than original ones as $\Delta H<\Delta H_{Shuff}$. $H(2)$ generally can not be distinguished from 0.5 thus long-range correlations have been destroyed by shuffling and shuffled series from the MSM model seem to capture the scaling behavior of the shuffled empirical data closer than the original ones. Still, shuffled MSM data show stronger multifractality than original ones in terms of differences of the scaling exponents.

Results of absolute values of returns in Table \ref{abs} show similar behaviors.
While shuffling destroys strong auto-correlations and $H(2)$ values drop from around 0.7-0.8 to 0.5, empirical data again show stronger degree of multifractality. 
Simulated series from MSM model show the same behavior. 
Although the difference is less than 0.1 in many cases (statistically insignificant), in lot of cases the difference is larger, even around 0.3. 
Table \ref{sqr} shows the results from the same analysis done on the squared returns, $\sum r_{t'}^2$. These results are again very similar but the degree of multifractality of original as well as shuffled series is even larger.

\begin{table}
\scriptsize
\caption{Differences $\Delta H$ and $\Delta H_{Shuff}$ on original and randomly shuffled data with different stochastic variables, $p_t$, $\sum |r_{t'}|$ and $\sum r_{t'}^2$. \textit{Emp.} refers to the empirical exponent values, $k=5$, $k=10$, $k=15$ and $k=20$ refer to the mean and standard deviation of the exponent values based on the 1000 simulated time series. }
\centering
\ra{1.2}
\begin{tabular}{lrrrrrrrrrrr}
\toprule
\multicolumn{12}{c}{$p_t$} \\
\hline
&  \multicolumn{5}{c}{$\Delta H$} & &  \multicolumn{5}{c}{$\Delta H_{Shuff}$} \\
\cmidrule{2-6} \cmidrule{8-12}
&Emp. & $k=5$ & $k=10$ & $k=15$ & $k=20$ &  &Emp. & $k=5$ & $k=10$ & $k=15$ & $k=20$  \\
\hline
Dow &0.059 & 0.031 & 0.026 & 0.027 & 0.027 &    &0.123 & 0.053 & 0.089 & 0.100 & 0.078 \\
Nik  &0.040 & 0.047 & 0.050 & 0.051 & 0.053 &    &0.086 & 0.088 & 0.170 & 0.203 & 0.145 \\
UK &0.048 & 0.015 & 0.011 & 0.011 & 0.011 &    &0.067 & 0.033 & 0.031 & 0.030 & 0.038 \\
Aud &0.039 & 0.041 & 0.034 & 0.033 & 0.034 &    &0.083 & 0.072 & 0.098 & 0.096 & 0.116 \\
TB1 &0.154 & 0.094 & 0.098 & 0.098 & 0.099 &    &0.246 & 0.181 & 0.277 & 0.281 & 0.294 \\
TB2 &0.165 & 0.121 & 0.125 & 0.130 & 0.129 &    &0.230 & 0.222 & 0.301 & 0.371 & 0.291 \\
TB3 &0.134 & 0.091 & 0.095 & 0.098 & 0.098 &    &0.199 & 0.168 & 0.320 & 0.359 & 0.274 \\
TB5 &0.102 & 0.103 & 0.109 & 0.111 & 0.109 &    &0.162 & 0.198 & 0.299 & 0.319 & 0.275 \\
TB10 &0.123 & 0.114 & 0.123 & 0.129 & 0.127 &    &0.151 & 0.217 & 0.273 & 0.406 & 0.352 \\
  \hline
\multicolumn{12}{c}{$\sum |r_{t'}|$} \\
\hline
&  \multicolumn{5}{c}{$\Delta H$} & &  \multicolumn{5}{c}{$\Delta H_{Shuff}$} \\
\cmidrule{2-6} \cmidrule{8-12}

Dow &-0.009 & 0.063 & 0.066 & 0.072 & 0.073 &    &0.171 & 0.070 & 0.122 & 0.113 & 0.092 \\
Nik  &-0.008 & 0.086 & 0.102 & 0.108 & 0.108 &    &0.112 & 0.096 & 0.159 & 0.189 & 0.134 \\
UK &0.053 & 0.037 & 0.035 & 0.038 & 0.038 &    &0.084 & 0.046 & 0.047 & 0.044 & 0.061 \\
Aud &0.049 & 0.077 & 0.078 & 0.084 & 0.084 &    &0.095 & 0.084 & 0.108 & 0.110 & 0.131 \\
TB1 &0.115 & 0.138 & 0.150 & 0.157 & 0.157 &    &0.258 & 0.172 & 0.252 & 0.266 & 0.293 \\
TB2 &0.125 & 0.161 & 0.177 & 0.181 & 0.181 &    &0.241 & 0.210 & 0.269 & 0.335 & 0.232 \\
TB3 &0.064 & 0.134 & 0.148 & 0.153 & 0.154 &    &0.210 & 0.160 & 0.278 & 0.309 & 0.245 \\
TB5 &0.044 & 0.145 & 0.157 & 0.165 & 0.164 &    &0.172 & 0.192 & 0.272 & 0.291 & 0.248 \\
TB10 &0.096 & 0.156 & 0.171 & 0.182 & 0.180 &    &0.178 & 0.185 & 0.262 & 0.373 & 0.313 \\
  \hline
\multicolumn{12}{c}{$\sum r_{t'}^2$ } \\
\hline
&  \multicolumn{5}{c}{$\Delta H$} & &  \multicolumn{5}{c}{$\Delta H_{Shuff}$} \\
\cmidrule{2-6} \cmidrule{8-12}
Dow &0.261 & 0.237 & 0.242 & 0.251 & 0.253 &    &0.387 & 0.251 & 0.343 & 0.335 & 0.291 \\
Nik  &0.084 & 0.274 & 0.302 & 0.310 & 0.311 &    &0.322 & 0.289 & 0.359 & 0.421 & 0.371 \\
UK &0.270 & 0.189 & 0.183 & 0.188 & 0.189 &    &0.285 & 0.221 & 0.228 & 0.205 & 0.240 \\
Aud &0.231 & 0.261 & 0.263 & 0.274 & 0.269 &    &0.300 & 0.289 & 0.315 & 0.331 & 0.342 \\
TB1 &0.383 & 0.349 & 0.371 & 0.378 & 0.379 &    &0.477 & 0.386 & 0.471 & 0.482 & 0.503 \\
TB2 &0.406 & 0.375 & 0.403 & 0.401 & 0.404 &    &0.459 & 0.414 & 0.467 & 0.529 & 0.459 \\
TB3 &0.295 & 0.343 & 0.368 & 0.371 & 0.374 &    &0.436 & 0.371 & 0.484 & 0.523 & 0.457 \\
TB5 &0.262 & 0.357 & 0.379 & 0.387 & 0.384 &    &0.374 & 0.399 & 0.487 & 0.512 & 0.451 \\
TB10 &0.448 & 0.369 & 0.393 & 0.405 & 0.403 &    &0.398 & 0.417 & 0.474 & 0.566 & 0.519 \\
 \bottomrule
\end{tabular}
\label{diffs}
\end{table}

\subsection{Does all multifractality come from the distribution?}

Our results from the comparison between the scaling exponents of the original series and shuffled ones suggest that most of the multifractality is caused solely by the broad, fat-tailed, distribution of the returns.
Indeed if we look at Table \ref{diffs} that summarizes the results and brings the differences of scaling exponents, $\Delta H$ and $\Delta H_{Shuff}$, we can see that in most of the cases, $\Delta H<\Delta H_{Shuff}$. 
Although most of these cases is within standard errors and $\Delta H$ and $\Delta H_{Shuff}$ can not be statistically distinguished, in some cases the difference is statistically significant. This suggests that there might be a negative bias in the estimation of the degree of multifractality brought by unknown time-correlations. This result is valid for prices as well as absolute returns and squared returns for both empirical as well as simulated data from MSM.

In the next section, we use several different processes to try to explain this type of behavior and possibly find the explanation which will help us to better understand the source of multifractality in the financial markets.

\section{Robustness of the results}

In order to demonstrate that our results are robust, we have generated time series from fractional motion, random walk with L\'{e}vy distributed steps (L\'{e}vy flights) and autoregressive fractionally integrated moving average (ARFIMA) process with stable innovations allowing us to put short memory into the processes.

The $\Delta H$ for the Brownian motion is equal to zero as it is uni-fractal ($H(1)=H(2)=H(3)=0.5$) and shuffling does not change this result. 
The $\Delta H$ of fractional Brownian motion  will also be zero  ($H(1)=H(2)=H(3)=H$) and the shuffling will destroy the long-range dependence in this case, and still $\Delta H_{Shuff}$ should be zero.
On the other hand, when we turn to uncorrelated series with fat-tailed return distributions, the situation is changing.
\cite{Kantelhardt2002} discuss the case of L\'{e}vy distribution and show that expected value of $H(q)$ for $q>\alpha$ is $1/q$ and expected value of $H(q)$ for $q\le\alpha$ is $1/\alpha$.
In \cite{barunika}, we use GHE method and show on the large finite sample study that this behavior holds.
Finally, we borrow ARFIMA process of \cite{kokoszka} to combine both long-range dependence and heavy tails into single process. Moreover, we would like to show also the impact of the simple short memory processes on the multifractality estimation.

\subsection{Comparison to simulations from $\alpha$-stable distribution}

Stable distributions form a class of probability laws with appealing theoretical properties which describe well-known stylized facts such as skewness, excess kurtosis and heavy tails. Such distributions are described by four parameters: $\alpha,\beta,\gamma,\delta$, where:   $\alpha$ is the characteristic exponent and $0<\alpha \leq2\ $; $\beta$ is the skewness parameter and $-1 \leq \beta \leq 1$; $\gamma$ and $\delta$ are scale and location parameters, respectively.
The stable distributions have two tails that are asymptotically power laws.

The $\alpha$-stable distribution can be described by a characteristic function:
\begin{equation}
 \footnotesize
\label{eq2}
\phi(u)=
\left\{
\begin{array}{lr}
\exp (-\gamma^{\alpha}|u|^{\alpha}[1+i\beta(\tan\frac{\pi\alpha}{2})$sign$(u)(|\gamma u|^{1-\alpha}-1)]+i\delta u ) & \alpha \ne 1 \\
\exp(-\gamma |u| [1+i\beta\frac{2}{\pi}$sign$(u)\ln (\gamma |u|)]+i\delta u ) & \alpha = 1\;\;,
\end{array}
\right.\end{equation}
which is the inverse Fourier transform of the probability density function \citep{Nolan2003}, i.e., $\phi(u)=E[\exp(iuX)]$.
In order to simulate random stable variables, we use a method of \cite{Chambers1976}. 
For all values of the parameters $\alpha<2$ and $-1<\beta<1$, we set the parameters $(\alpha,\beta,\gamma,\delta)$  to $(\alpha,0,{\sqrt{2}}/{2},0)$, where $\alpha=\{1.2,1.4,1.6,1.8,2\}$.
We choose parameters with special case of $\alpha=2$ being Gaussian distribution, so the lower the $\alpha$, the fatter the tails other parameters being equal. This will allow us to show the direct impact of the fat tails on the multi-scaling.

\begin{table}
\scriptsize
\caption{$H(1)$, $H(2)$, $H(3)$ and $\Delta H$ of simulated random variables from $\alpha$-stable distribution with different $\alpha$. Values refer to the mean and standard deviation of the exponent values based on the 1000 simulated time series. Note that for $\alpha=2$, we have Gaussian random variables with mean zero and variance one. }
\centering
\ra{1.2}
\begin{tabular}{lrrrrrrrrrrr}

\toprule
\multicolumn{12}{c}{\bf{O R I G I N A L}} \\
\hline
&  \multicolumn{2}{c}{$H(1)$} & & \multicolumn{2}{c}{$H(2)$} & & \multicolumn{2}{c}{$H(3)$} & & \multicolumn{2}{c}{$\Delta H$} \\
\cmidrule{2-3} \cmidrule{5-6} \cmidrule{8-9} \cmidrule{11-12}
$\alpha=1.2$ &0.811 & (0.059) &    &0.499 & (0.006) &    &0.333 & (0.007) & & \bf{0.478} \\
$\alpha=1.4$ &0.705 & (0.035) &    &0.499 & (0.014) &    &0.334 & (0.020)  & & \bf{0.371} \\
$\alpha=1.6$ &0.626 & (0.040) &    &0.500 & (0.007) &    &0.340 & (0.013) & & \bf{0.286} \\
$\alpha=1.8$ &0.554 & (0.017) &    &0.499 & (0.009) &    &0.363 & (0.026) & & \bf{0.191} \\
$\alpha=2$ &0.499 & (0.007) &    &0.499 & (0.007) &    &0.499 & (0.008) & & \bf{0.000} \\
\hline
\multicolumn{12}{c}{\bf{S H U F F L E D}} \\
\hline
&  \multicolumn{2}{c}{$H(1)$} & & \multicolumn{2}{c}{$H(2)$} & & \multicolumn{2}{c}{$H(3)$} & & \multicolumn{2}{c}{$\Delta H$} \\
\cmidrule{2-3} \cmidrule{5-6} \cmidrule{8-9} \cmidrule{11-12}
$\alpha=1.2$ &0.811 & (0.058) &    &0.500 & (0.004) &    &0.334 & (0.005) & & \bf{0.477} \\
$\alpha=1.4$ &0.704 & (0.036) &    &0.499 & (0.006) &    &0.334 & (0.008) & & \bf{0.369} \\
$\alpha=1.6$ &0.624 & (0.036) &    &0.499 & (0.006) &    &0.340 & (0.012) & & \bf{0.285} \\
$\alpha=1.8$ &0.555 & (0.019) &    &0.500 & (0.008) &    &0.363 & (0.026) & & \bf{0.192} \\
$\alpha=2$ &0.500 & (0.010) &    &0.499 & (0.009) &    &0.499 & (0.009) & & \bf{0.001}  \\
 \bottomrule
\end{tabular}
\label{stable}
\end{table}
From Table \ref{stable}, we can see that theoretical expectations hold. For the example when $\alpha=1.6$ corresponding to the estimates on empirical data \citep{barunikb}, we have the expected value of $H(1)=1/\alpha=0.625$, and expected value of $H(3)=1/3\simeq 0.33$ and thus expected value of $\Delta H=\Delta H_{Shuff}=0.292$ approximately, which are all in good agreement with what reported  in  Table \ref{stable} within the standard errors of GHE estimator. 
In general, Table \ref{stable} confirms that these relations are well followed and the Hurst exponents (as well as the differences) are the same for original and shuffled data.
Thus we can see that heavy tails of the distribution can directly cause large difference in the $H(q)$ and spurious multifractality on the finite sample data and, because the data have no correlation structure, this multifractality can not be destroyed by shuffling.

Figure \ref{plotHqstable} (b) shows also scaling functions of the simulated $\alpha$-stable random variables with different heavy tails. 
Gaussian case with $\alpha=2$ shows straight line as expected, as it is unifractal and $H(1)=H(2)=H(3)=0.5$. 
But heavier the tails are (the lower the $\alpha$ is), stronger the multi-scaling is. 
Again, shuffled data shows no statistically significant difference as expected. 
The case of $\alpha=1.8$ replicates well the behavior of the data studied in the previous sections.

\subsection{Comparison with simulations from fractional Brownian motion}

We here perform the same analysis with the data simulated from fractional Brownian motion \citep{mandelbrotvanness,beran} with different long range parameters $H=\{0.3,0.4,0.5,0.6,0.7\}$.
The $\Delta H$ of fractional Brownian motion with Hurst exponent equal to $H$ is zero as this is a uni-fractal process and $H(1)=H(2)=H(3)=H$.
Shuffling will destroy the long-range dependence and the exponents must become $H(1)=H(2)=H(3)=0.5$, while $\Delta H_{Shuff}$ is zero.

\begin{table}
\scriptsize
\caption{$H(1)$, $H(2)$, $H(3)$ and $\Delta H$ of simulated random variables from fractional Brownian motion fGn($d$) with different memory parameter $d$. Values refer to the mean and standard deviation of the exponent values based on the 1000 simulated time series. Note that for memory parameter $d=0.5$, we have Gaussian random variables with mean zero and variance one.}
\centering
\ra{1.2}
\begin{tabular}{lrrrrrrrrrrr}
\toprule
\multicolumn{12}{c}{\bf{O R I G I N A L}} \\
\hline
&  \multicolumn{2}{c}{$H(1)$} & & \multicolumn{2}{c}{$H(2)$} & & \multicolumn{2}{c}{$H(3)$} & & \multicolumn{2}{c}{$\Delta H$} \\
\cmidrule{2-3} \cmidrule{5-6} \cmidrule{8-9} \cmidrule{11-12}
fGn(0.3) &0.300 & (0.008) &    &0.299 & (0.008) &    &0.299 & (0.009) & & \bf{0.000}\\
fGn(0.4) &0.400 & (0.008) &    &0.400 & (0.008) &    &0.399 & (0.008) & &  \bf{0.000}\\
fGn(0.5) &0.501 & (0.009) &    &0.500 & (0.009) &    &0.499 & (0.010) & &  \bf{0.000}\\
fGn(0.6) &0.598 & (0.010) &    &0.598 & (0.009) &    &0.598 & (0.009) & &  \bf{0.000}\\
fGn(0.7) &0.697 & (0.011) &    &0.696 & (0.010) &    &0.696 & (0.010) & &  \bf{0.000}\\
\hline
\multicolumn{12}{c}{\bf{S H U F F L E D}} \\
\hline
&  \multicolumn{2}{c}{$H(1)$} & & \multicolumn{2}{c}{$H(2)$} & & \multicolumn{2}{c}{$H(3)$} & & \multicolumn{2}{c}{$\Delta H$} \\
\cmidrule{2-3} \cmidrule{5-6} \cmidrule{8-9} \cmidrule{11-12}

fGn(0.3) &0.499 & (0.009) &    &0.499 & (0.009) &    &0.499 & (0.009) & &  \bf{0.000}\\
fGn(0.4) &0.498 & (0.008) &    &0.498 & (0.008) &    &0.497 & (0.008) & &  \bf{0.000}\\
fGn(0.5) &0.500 & (0.008) &    &0.499 & (0.008) &    &0.499 & (0.008) & &  \bf{0.000}\\
fGn(0.6) &0.500 & (0.008) &    &0.499 & (0.008) &    &0.499 & (0.009) & &  \bf{0.000}\\
fGn(0.7) &0.499 & (0.009) &    &0.499 & (0.009) &    &0.498 & (0.009) & &  \bf{0.000}\\
 \bottomrule
\end{tabular}
\label{fgn}
\end{table}

\begin{figure}
   \centering
   \includegraphics[width=5in]{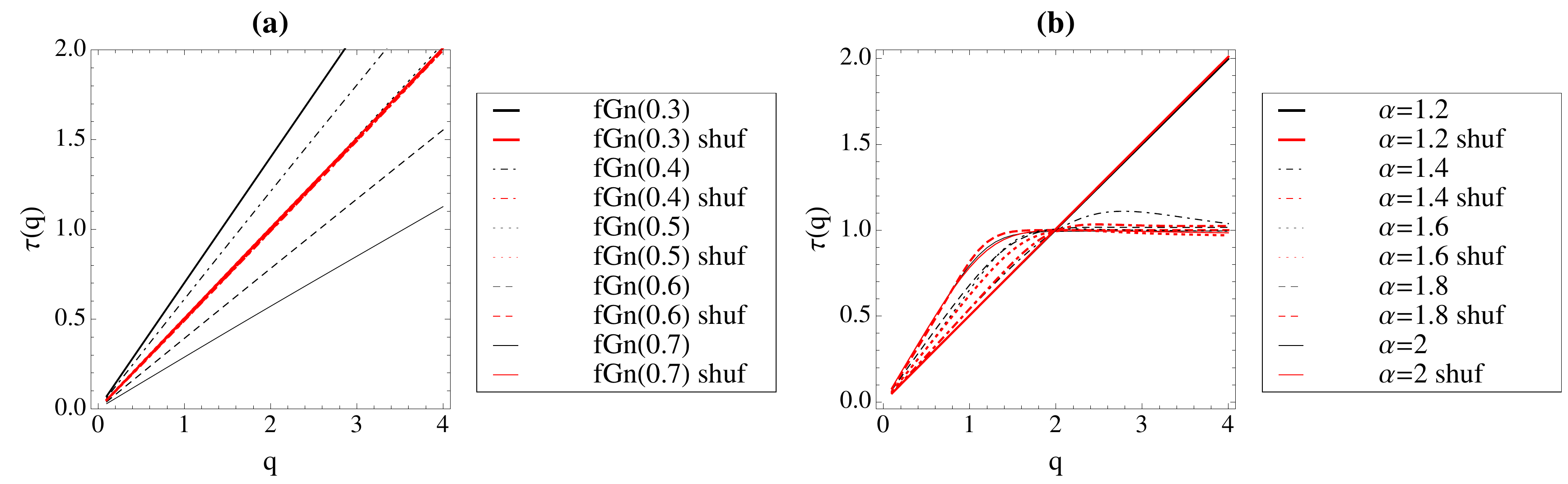}
   \caption{(Color online) (a) Scaling functions $\tau(q)=qH(q)$ for simulated data from fractional Brownian motion for different values of long range dependence. Shuffled series in red. (b) Scaling functions $\tau(q)=qH(q)$ for simulated data from $\alpha$-stable distributions.}
   \label{plotHqstable}
\end{figure}

Table \ref{fgn} summarises the results. We can see that GHE method is able to estimate all values of Hurst parameter on simulated series very precisely, and the result is as expected. 
As the fractional Brownian motion is uni-fractal, $\Delta H$ is equal to zero as $H(1)=H(2)=H(3)=H$. 
When we shuffle the series, all long-range dependence is destroyed and $H(1)=H(2)=H(3)=0.5$.

Figure \ref{plotHqstable} (a) shows also scaling functions of the simulated fractional Brownian motion with different long memory parameters.
All cases show straight line consistently with the fact that  all the simulated data are uni-fractal.
Moreover, shuffled series are the same as the Gaussian case with $H=0.5$.

\subsection{Comparison with simulations from ARFIMA process with stable innovations}

While previous two sections allowed us to compare our results to the long-range and heavy tailed processes separately, we would like to complete the analysis by comparing the results to the heavy tailed process, which exhibits asymptotic self-similarity and/or long-range dependence. 
Moreover, we would like to study how simple short memory dependence influences the multifractality measures. 
For this purpose, we use  autoregressive fractionally integrated moving average (ARFIMA) process with $\alpha$-stable innovations studied in \cite{kokoszka}. 

These models, denoted ARFIMA$(p,d,q)$, $p,q\in \mathbb{N}$ are extensions to linear ARIMA$(p,d,q)$ models, replacing the integer differencing exponent $d$ with an arbitrary fractional real number $0<d<1-1/\alpha$ and $1<\alpha<2$ heavy tails parameter from $\alpha$-stable innovations. A fractional ARIMA process $Y=\{Y(k),k\in \mathbb{Z}\}$ with $\alpha$-stable innovations is defined as the stationary solution to the back-shift operator equation
\begin{equation}
\Phi_p(B)Y(k)=\Theta_q(B)(I-B)^{-d}Z_{\alpha}(k), k\in\mathbb{Z},
\end{equation}
where the innovations $Z_{\alpha}$ are an \textit{i.i.d.} standard $\alpha$-stable random variables, $BY(k):=Y(k-1)$ and $\Phi_p(z)=1-\phi_1z-\phi_2z^2-\ldots-\phi_pz^p$, and $\Theta_q(z)=1-\theta_1z-\theta_2z^2-\ldots-\theta_qz^q$ with roots outside the unit disk $\{z\in\mathbb{C},L|z|\le1\}$. For more details about this process, see \cite{kokoszka}.

\begin{landscape}
\begin{table}
\tiny
\ra{1.2}
\caption{$H(1)$, $H(2)$, $H(3)$ and $\Delta H$ of simulated random variables from fractional autoregressive moving average, ARFIMA$(0,d,0)$ process with $\alpha$-stable innovations. Values refer to the mean and standard deviation of the exponent values based on the 1000 simulated time series. Note that for memory parameter $d=0$ with tail $\alpha=2$, we have Gaussian random variables with mean zero and variance one.}\centering
\begin{tabular}{rlrrrrrrrrrrrrrrrrrrrrr}
\toprule
& & \multicolumn{10}{c}{\bf{O R I G I N A L}} & & \multicolumn{10}{c}{\bf{S H U F F L E D}} \\
\cmidrule{3-12} \cmidrule{14-23}
& & \multicolumn{2}{c}{H(1)} & & \multicolumn{2}{c}{H(2)} & & \multicolumn{2}{c}{H(3)} & & \multicolumn{1}{c}{$\Delta H$} & & \multicolumn{2}{c}{H(1)} & & \multicolumn{2}{c}{H(2)} & & \multicolumn{2}{c}{H(3)} & & \multicolumn{1}{c}{$\Delta H$} \\
\cmidrule{3-4} \cmidrule{6-7} \cmidrule{9-10} \cmidrule{12-12}  \cmidrule{14-15}  \cmidrule{17-18}  \cmidrule{20-21}  \cmidrule{23-23}
\multirow{5}{*}{\rotatebox{90}{$\alpha=1.2$}} & $d=-0.2$ &0.622 & (0.045) &    &0.334 & (0.024) &    &0.165 & (0.023) &  & \bf{-0.457} && 0.789 & (0.051) &    &0.498 & (0.010) &    &0.332 & (0.014) &  & \bf{0.457} \\ 
&$d=-0.1$ &0.714 & (0.045) &    &0.415 & (0.011) &    &0.248 & (0.014) &  & \bf{-0.466}  && 0.796 & (0.050) &    &0.500 & (0.007) &    &0.334 & (0.010) &  & \bf{0.462} \\ 
&$d=0$ &0.810 & (0.053) &    &0.500 & (0.005) &    &0.333 & (0.007) &  & \bf{-0.477}  && 0.809 & (0.051) &    &0.498 & (0.014) &    &0.332 & (0.020) &  & \bf{0.477} \\ 
&$d=0.1$  &0.872 & (0.044) &    &0.587 & (0.011) &    &0.419 & (0.014) &  & \bf{-0.453}  && 0.780 & (0.045) &    &0.500 & (0.005) &    &0.334 & (0.005) &  & \bf{0.446} \\ 
&$d=0.2$ &0.921 & (0.032) &    &0.679 & (0.013) &    &0.509 & (0.019) &  & \bf{-0.412}  && 0.741 & (0.039) &    &0.499 & (0.006) &    &0.335 & (0.009) &  & \bf{0.406} \\ 
& & \\
\multirow{5}{*}{\rotatebox{90}{$\alpha=1.4$}} & $d=-0.2$ &0.529 & (0.031) &    &0.336 & (0.017) &    &0.170 & (0.022) &  & \bf{-0.359}  && 0.689 & (0.030) &    &0.500 & (0.010) &    &0.336 & (0.015) &  & \bf{0.353} \\ 
&$d=-0.1$&0.616 & (0.036) &    &0.416 & (0.011) &    &0.250 & (0.013) &  & \bf{-0.365}  && 0.698 & (0.036) &    &0.499 & (0.006) &    &0.335 & (0.008) &  & \bf{0.363} \\ 
&$d=0$&0.705 & (0.041) &    &0.498 & (0.012) &    &0.332 & (0.014) &  & \bf{-0.373}  && 0.705 & (0.042) &    &0.500 & (0.008) &    &0.336 & (0.011) &  & \bf{0.369} \\ 
&$d=0.1$ &0.799 & (0.041) &    &0.588 & (0.009) &    &0.421 & (0.014) &  & \bf{-0.378}  && 0.707 & (0.042) &    &0.499 & (0.006) &    &0.335 & (0.009) &  & \bf{0.372} \\ 
&$d=0.2$ &0.867 & (0.038) &    &0.684 & (0.031) &    &0.516 & (0.048) &  & \bf{-0.351}  && 0.685 & (0.043) &    &0.499 & (0.007) &    &0.335 & (0.011) &  & \bf{0.350} \\ 
& & \\
\multirow{5}{*}{\rotatebox{90}{$\alpha=1.6$}} & $d=-0.2$ &0.455 & (0.024) &    &0.336 & (0.019) &    &0.176 & (0.029) &  & \bf{-0.279}  && 0.618 & (0.024) &    &0.500 & (0.007) &    &0.343 & (0.015) &  & \bf{0.275} \\ 
&$d=-0.1$&0.544 & (0.041) &    &0.417 & (0.011) &    &0.256 & (0.021) &  & \bf{-0.288}  && 0.624 & (0.036) &    &0.500 & (0.007) &    &0.341 & (0.013) &  & \bf{0.283} \\ 
&$d=0$ &0.623 & (0.025) &    &0.501 & (0.011) &    &0.342 & (0.020) &  & \bf{-0.281}  && 0.623 & (0.024) &    &0.500 & (0.008) &    &0.341 & (0.013) &  & \bf{0.282} \\ 
&$d=0.1$  &0.708 & (0.026) &    &0.587 & (0.009) &    &0.428 & (0.017) &  & \bf{-0.280}  && 0.618 & (0.025) &    &0.499 & (0.007) &    &0.341 & (0.015) &  & \bf{0.277} \\ 
&$d=0.2$&0.795 & (0.031) &    &0.678 & (0.012) &    &0.515 & (0.023) &  & \bf{-0.279}  &&  0.614 & (0.033) &    &0.501 & (0.008) &    &0.344 & (0.016) &  & \bf{0.270} \\ 
& & \\
\multirow{5}{*}{\rotatebox{90}{$\alpha=1.8$}} & $d=-0.2$ &0.396 & (0.020) &    &0.341 & (0.014) &    &0.200 & (0.035) &  & \bf{-0.196}  &&  0.554 & (0.020) &    &0.499 & (0.009) &    &0.361 & (0.025) &  & \bf{0.194} \\ 
&$d=-0.1$&0.472 & (0.022) &    &0.417 & (0.010) &    &0.281 & (0.030) &  & \bf{-0.191}  && 0.553 & (0.020) &    &0.500 & (0.008) &    &0.364 & (0.030) &  & \bf{0.189} \\ 
&$d=0$&0.558 & (0.027) &    &0.499 & (0.008) &    &0.360 & (0.026) &  & \bf{-0.198}  && 0.557 & (0.024) &    &0.499 & (0.008) &    &0.360 & (0.027) &  & \bf{0.198} \\ 
&$d=0.1$  &0.645 & (0.016) &    &0.586 & (0.008) &    &0.443 & (0.025) &  & \bf{-0.202}  && 0.556 & (0.016) &    &0.501 & (0.009) &    &0.361 & (0.027) &  & \bf{0.195} \\ 
&$d=0.2$ &0.730 & (0.015) &    &0.680 & (0.009) &    &0.547 & (0.033) &  & \bf{-0.183}  && 0.550 & (0.014) &    &0.499 & (0.018) &    &0.368 & (0.051) &  & \bf{0.182} \\ 
& & \\
\multirow{5}{*}{\rotatebox{90}{$\alpha=2$}} & $d=-0.2$ &0.340 & (0.010) &    &0.340 & (0.009) &    &0.340 & (0.009) &  & \bf{-0.000}  && 0.500 & (0.009) &    &0.500 & (0.009) &    &0.500 & (0.009) &  & \bf{0.000} \\ 
&$d=-0.1$&0.416 & (0.008) &    &0.416 & (0.008) &    &0.416 & (0.008) &  & \bf{-0.000}  && 0.499 & (0.009) &    &0.499 & (0.009) &    &0.499 & (0.009) &  & \bf{0.000} \\ 
&$d=0$ &0.500 & (0.010) &    &0.500 & (0.009) &    &0.500 & (0.009) &  & \bf{-0.000}  && 0.499 & (0.008) &    &0.499 & (0.008) &    &0.499 & (0.008) &  & \bf{0.001} \\ 
&$d=0.1$  &0.589 & (0.010) &    &0.588 & (0.009) &    &0.587 & (0.009) &  & \bf{-0.002} &&  0.501 & (0.009) &    &0.501 & (0.009) &    &0.501 & (0.009) &  & \bf{0.000} \\ 
&$d=0.2$ &0.680 & (0.011) &    &0.680 & (0.011) &    &0.680 & (0.011) &  & \bf{-0.001}  && 0.500 & (0.009) &    &0.500 & (0.009) &    &0.500 & (0.009) &  & \bf{0.000} \\ 
 \bottomrule
\end{tabular}
\label{farima1}
\end{table}
\end{landscape}

\begin{landscape}
\begin{table}
\tiny
\ra{1.2}
\caption{$H(1)$, $H(2)$, $H(3)$ and $\Delta H$ of simulated random variables from fractional autoregressive moving average, ARFIMA$(1,d,0)$ process with $\alpha$-stable innovations and AR(1) parameter equal to 0.4. Values refer to the mean and standard deviation of the exponent values based on the 1000 simulated time series. Note that for memory parameter $d=0$ with tail $\alpha=2$, we have Gaussian random variables with mean zero and variance one.}
\centering
\begin{tabular}{rlrrrrrrrrrrrrrrrrrrrrr}
\toprule
& & \multicolumn{10}{c}{\bf{O R I G I N A L}} & & \multicolumn{10}{c}{\bf{S H U F F L E D}} \\
\cmidrule{3-12} \cmidrule{14-23}
& & \multicolumn{2}{c}{H(1)} & & \multicolumn{2}{c}{H(2)} & & \multicolumn{2}{c}{H(3)} & & \multicolumn{1}{c}{$\Delta H$} & & \multicolumn{2}{c}{H(1)} & & \multicolumn{2}{c}{H(2)} & & \multicolumn{2}{c}{H(3)} & & \multicolumn{1}{c}{$\Delta H$} \\
\cmidrule{3-4} \cmidrule{6-7} \cmidrule{9-10} \cmidrule{12-12}  \cmidrule{14-15}  \cmidrule{17-18}  \cmidrule{20-21}  \cmidrule{23-23}
\multirow{5}{*}{\rotatebox{90}{$\alpha=1.2$}} & $d=-0.2$ &0.786 & (0.042) &    &0.508 & (0.013) &    &0.341 & (0.016) &  & \bf{-0.445} & &  0.790 & (0.058) &    &0.501 & (0.014) &    &0.335 & (0.019) &  & \bf{0.456} \\ 
&$d=-0.1$ &0.831 & (0.034) &    &0.579 & (0.013) &    &0.426 & (0.019) &  & \bf{-0.405} & & 0.787 & (0.048) &    &0.499 & (0.005) &    &0.333 & (0.007) &  & \bf{0.454} \\   
&$d=0$ &0.877 & (0.032) &    &0.648 & (0.016) &    &0.506 & (0.025) &  & \bf{-0.370} & & 0.802 & (0.054) &    &0.499 & (0.009) &    &0.332 & (0.011) &  & \bf{0.470} \\ 
&$d=0.1$ &0.924 & (0.031) &    &0.720 & (0.014) &    &0.587 & (0.022) &  & \bf{-0.337} & & 0.787 & (0.055) &    &0.499 & (0.006) &    &0.333 & (0.007) &  & \bf{0.453} \\ 
&$d=0.2$ &0.945 & (0.022) &    &0.789 & (0.013) &    &0.665 & (0.023) &  & \bf{-0.280} & & 0.731 & (0.045) &    &0.502 & (0.008) &    &0.338 & (0.011) &  & \bf{0.393} \\ 
& & \\
\multirow{5}{*}{\rotatebox{90}{$\alpha=1.4$}} &$d=-0.2$ &0.706 & (0.044) &    &0.510 & (0.014) &    &0.343 & (0.022) &  & \bf{-0.362} & & 0.699 & (0.054) &    &0.500 & (0.006) &    &0.336 & (0.008) &  & \bf{0.364} \\ 
&$d=-0.1$ &0.759 & (0.040) &    &0.576 & (0.011) &    &0.423 & (0.016) &  & \bf{-0.336} & & 0.706 & (0.059) &    &0.500 & (0.006) &    &0.335 & (0.009) &  & \bf{0.371} \\ 
&$d=0$ &0.806 & (0.032) &    &0.650 & (0.012) &    &0.510 & (0.021) &  & \bf{-0.296}  & & 0.703 & (0.054) &    &0.495 & (0.039) &    &0.332 & (0.043) &  & \bf{0.371} \\ 
&$d=0.1$ &0.864 & (0.033) &    &0.717 & (0.015) &    &0.585 & (0.026) &  & \bf{-0.279}  & & 0.698 & (0.056) &    &0.501 & (0.009) &    &0.340 & (0.015) &  & \bf{0.358} \\ 
&$d=0.2$ &0.911 & (0.023) &    &0.790 & (0.012) &    &0.669 & (0.023) &  & \bf{-0.242}  & & 0.677 & (0.046) &    &0.500 & (0.008) &    &0.339 & (0.013) &  & \bf{0.338} \\ 
& & \\
\multirow{5}{*}{\rotatebox{90}{$\alpha=1.6$}} & $d=-0.2$ &0.633 & (0.033) &    &0.509 & (0.011) &    &0.349 & (0.020) &  & \bf{-0.284}  & & 0.625 & (0.042) &    &0.497 & (0.011) &    &0.337 & (0.019) &  & \bf{0.288} \\ 
&$d=-0.1$ &0.687 & (0.027) &    &0.580 & (0.011) &    &0.431 & (0.016) &  & \bf{-0.256}  & & 0.621 & (0.035) &    &0.499 & (0.008) &    &0.341 & (0.015) &  & \bf{0.280} \\ 
&$d=0$ &0.746 & (0.025) &    &0.649 & (0.013) &    &0.511 & (0.024) &  & \bf{-0.234}  & & 0.624 & (0.046) &    &0.499 & (0.009) &    &0.342 & (0.017) &  & \bf{0.281} \\ 
&$d=0.1$ &0.809 & (0.024) &    &0.722 & (0.012) &    &0.598 & (0.024) &  & \bf{-0.211}  & & 0.619 & (0.042) &    &0.499 & (0.008) &    &0.345 & (0.018) &  & \bf{0.274} \\ 
&$d=0.2$ &0.868 & (0.018) &    &0.788 & (0.011) &    &0.671 & (0.022) &  & \bf{-0.197}  & & 0.613 & (0.033) &    &0.500 & (0.009) &    &0.349 & (0.020) &  & \bf{0.264} \\ 
& & \\
\multirow{5}{*}{\rotatebox{90}{$\alpha=1.8$}} & $d=-0.2$ &0.566 & (0.022) &    &0.512 & (0.008) &    &0.377 & (0.029) &  & \bf{-0.189}  & & 0.555 & (0.024) &    &0.500 & (0.009) &    &0.366 & (0.028) &  & \bf{0.189} \\ 
&$d=-0.1$ &0.628 & (0.015) &    &0.579 & (0.008) &    &0.456 & (0.027) &  & \bf{-0.172}  & & 0.553 & (0.019) &    &0.499 & (0.009) &    &0.367 & (0.030) &  & \bf{0.186} \\ 
&$d=0$ &0.692 & (0.013) &    &0.649 & (0.010) &    &0.536 & (0.033) &  & \bf{-0.156}  & & 0.552 & (0.019) &    &0.500 & (0.008) &    &0.373 & (0.031) &  & \bf{0.178} \\ 
&$d=0.1$ &0.761 & (0.018) &    &0.719 & (0.010) &    &0.611 & (0.032) &  & \bf{-0.149}  & & 0.555 & (0.025) &    &0.500 & (0.008) &    &0.373 & (0.034) &  & \bf{0.182} \\ 
&$d=0.2$ &0.826 & (0.015) &    &0.789 & (0.009) &    &0.695 & (0.028) &  & \bf{-0.131}  & & 0.549 & (0.025) &    &0.499 & (0.008) &    &0.379 & (0.033) &  & \bf{0.170} \\ 
& & \\
\multirow{5}{*}{\rotatebox{90}{$\alpha=2$}} & $d=-0.2$ &0.510 & (0.008) &    &0.510 & (0.008) &    &0.510 & (0.008) &  & \bf{-0.000}  & & 0.500 & (0.008) &    &0.501 & (0.008) &    &0.501 & (0.008) &  & \bf{0.000} \\ 
&$d=-0.1$ &0.581 & (0.007) &    &0.580 & (0.007) &    &0.580 & (0.008) &  & \bf{-0.001}  & & 0.501 & (0.009) &    &0.500 & (0.009) &    &0.500 & (0.009) &  & \bf{0.001} \\ 
&$d=0$ &0.649 & (0.008) &    &0.649 & (0.008) &    &0.648 & (0.009) &  & \bf{-0.000}  & & 0.499 & (0.008) &    &0.500 & (0.009) &    &0.500 & (0.010) &  & \bf{0.000} \\ 
&$d=0.1$ &0.719 & (0.009) &    &0.719 & (0.008) &    &0.720 & (0.008) &  & \bf{0.001}  & & 0.499 & (0.008) &    &0.499 & (0.008) &    &0.499 & (0.008) &  & \bf{0.000} \\ 
&$d=0.2$ &0.790 & (0.010) &    &0.790 & (0.009) &    &0.789 & (0.009) &  & \bf{-0.001}  & & 0.500 & (0.009) &    &0.500 & (0.008) &    &0.500 & (0.008) &  & \bf{0.000} \\  
 \bottomrule
\end{tabular}
\label{farima2}
\end{table}
\end{landscape}

To obtain the realizations from ARFIMA process, we use the algorithm proposed by \cite{stoev}. We simulate the ARFIMA process for all combinations of different tails $\alpha=\{1.2,1.4,1.6,1.8,2\}$ and long range parameter $d=\{-0.2,-0.1,0,0.1,0.2\}$. The usual $H$ parameter we work with through this text translates into $H=d+1/\alpha$ in this process. First, we simulate ARFIMA$(0,d,0)$, hence we obtain heavy tailed process which exhibits long range dependence. Second, we add simple AR(1) dependence to it and simulate ARFIMA$(1,d,0)$ with AR(1) parameter equal to $0.4$. Thus we put so-called short memory into the heavy tailed long memory process.

Tables \ref{farima1}, \ref{farima2} summarize the results for ARFIMA$(0,d,0)$ and ARFIMA$(1,d,0)$ respectively. Both tables report estimated $H(1),H(2)$ and $H(3)$ for all combinations as well as $H(1),H(2)$ and $H(3)$ for its shuffled counterparts together with the  differences measuring the degree of multifractality. We expect that shuffling will destroy any dependence structure in the data. Similarly to previous sections, we use 1000 realizations to obtain average values of estimates with standard deviations. 

The results suggest that GHE serves as precise method for estimation also in the case of heavy tail process exhibiting long-range dependence. When we shuffle the ARFIMA$(0,d,0)$ simulated time series, all long-range dependence is destroyed and $H(q)$ estimates are the same as those of pure $\alpha$-stable process reported in previous section. The multifractality of this process is solely caused by the fat-tailed distribution. This can be seen by comparison of the results to previous section, but also that all shuffled series with the same $\alpha$ parameter has the same properties.

In contrast, Table \ref{farima2} reporting the results from ARFIMA$(1,d,0)$ simulations reveals very interesting findings. Presence of short memory (AR(1)) process translates into the upward bias in the $H(q)$ estimates. Moreover, the difference $\Delta H$ is much lower in comparison to the same process without the short memory (significant in some cases). When we shuffle ARFIMA$(1,d,0)$, we arrive to the same results as in Table \ref{farima1} and in previous section which considered only pure $\alpha$-stable process. This means that we destroy all the dependence.

This final exercise brought us very interesting result, that short memory may bring significant contributions to the estimated degree of multifractality. 

\section{Conclusion}
In this paper, we present and discuss several new results on the multifractal behavior of different financial market datasets.
By computing and comparing the scaling exponents for empirical and simulated data, we show that the generalized Hurst exponent approach is a powerful tool to detect the scaling property of financial markets data for different stochastic variables, and it is also a good tool to test the reliability of models such as MSM.
By using Monte Carlo simulations as well as empirical analysis we show that the MSM model is able to capture the multi-scaling behavior of the volatility series, but it is not able to replicate the scaling properties of the prices. 
Large simulations show some agreement of the MSM model with empirical data, but most of the multifractality seem to be originated from the broad distribution of the series and not from the time-dependencies in the model. 
However, we observed a consistent increment of the multifractality after shuffling, i.e. when all temporal correlations are removed while the return distribution is preserved.
We demonstrated that in some cases the increment is significantly larger than the confidence interval. 
We argue that short memory time-correlation in the data may be the cause this increase in the multifractality.

We further compare our results to the simulated series from uni-fractal fractional Brownian motion containing only long-range dependence, as well as heavy-tailed $\alpha$-stable L\'evy distribution and  autoregressive fractionally integrated moving average process with stable innovations and the outcomes confirm our result.

In conclusion, from the present analysis we can argue that most of the multifractality observed in the stock markets data is the consequence to a broad fat-tailed distribution of the returns, while some part may be affected by the presence of short memory process in the data.
Our results are statistically robust and hold to various types of financial data; stock market indices, exchange rates as well as interest rates.

\section*{Acknowledgements}
Tiziana Di Matteo and Tomaso Aste acknowledge partial support by COST MP0801 project. Jozef Barunik acknowledges partial support from the Czech Science Foundation under Grants 402/09/0965 and 402/09/H045.

\bibliography{references_hurst}
\bibliographystyle{chicago}

\end{document}